\documentclass[aps,prl,twocolumn,superscriptaddress,showpacs,longbibliography]{revtex4-2}

\usepackage{amsmath, amssymb,wasysym,graphicx}
\usepackage{appendix}
\usepackage[ruled]{algorithm2e}

\usepackage[utf8]{inputenc}
\usepackage[T1]{fontenc}
\usepackage{xcolor}

\usepackage{diagbox}
\usepackage{tabularx}
\usepackage{gensymb}
\usepackage{siunitx}
\usepackage{ulem}

\IfFileExists{newtxtext.sty}
{\usepackage{newtxtext,newtxmath}}
{\IfFileExists{stix.sty}
	{\usepackage{stix}}
	{\IfFileExists{mathptmx.sty}
		{\usepackage{mathptmx}}{} } }

\usepackage{textcomp}

\usepackage{bm}
\usepackage{booktabs,siunitx}
\usepackage{dsfont}
\usepackage{color}
\definecolor{LinkColor}{rgb}{0.256,0.439,0.588}
\usepackage{hyperref}
\hypersetup{
	pdfauthor={good guys},
	pdftitle={good title},
	colorlinks=true,
	citecolor=LinkColor,
	linkcolor=LinkColor,
	urlcolor=LinkColor
}

\def\bk{{\bm{k}}}
\def\bK{{\mathrm{K}}}

\def\br{{\bm{r}}}
\def\bq{{\bm{q}}}
\def\bQ{{\bm{Q}}}
\def\bG{{\bm{G}}}
\def\mQ{{\mathcal{Q}}}
\def\G{{\mathbf{G}}}
\def\i{{\mathrm{i}}}
\def\I{{\mathrm{I}}}
\def\e{{\mathrm{e}}}
\def\c{{\mathrm{c}}}
\def\det{{\mathrm{det}}}
\def\Tr{{\mathrm{Tr}}}
\def\vd{\overrightarrow{d}}
\def\vc{\overrightarrow{c}}

\usepackage{multirow}

\begin{document}
\title{Angle-Tuned Gross-Neveu Quantum Criticality in Twisted Bilayer Graphene: \\ A Quantum Monte Carlo Study}   
\author{Cheng Huang}
\affiliation{Department of Physics and HK Institute of Quantum Science \& Technology, The University of Hong Kong, Pokfulam Road, Hong Kong SAR, China}

\author{Nikolaos Parthenios} 
\affiliation{Max Planck Institute for Solid State Reserach, Heisenbergstr. 1, 70569 Stuttgart Germany}
\affiliation{School of Natural Sciences, Technische Universit\"at München, 85748 Garching, Germany}

\author{Maksim Ulybyshev}
\affiliation{Institut f\"ur Theoretische Physik und Astrophysik, Universit\"at W\"urzburg, 97074 W\"urzburg, Germany}

\author{Xu Zhang}
\affiliation{Department of Physics and HK Institute of Quantum Science \& Technology, The University of Hong Kong, Pokfulam Road, Hong Kong SAR, China}
\affiliation{Department of Physics and Astronomy, Ghent University, Krijgslaan 281, 9000 Gent, Belgium}

\author{Fakher F. Assaad}
\affiliation{Institut f\"ur Theoretische Physik und Astrophysik, Universit\"at W\"urzburg, 97074 W\"urzburg, Germany}
\affiliation{W\"urzburg-Dresden Cluster of Excellence ct.qmat, Am Hubland, 97074 W\"urzburg, Germany}

\author{Laura Classen}
\email{l.classen@fkf.mpg.de}
\affiliation{Max Planck Institute for Solid State Reserach, Heisenbergstr. 1, 70569 Stuttgart Germany}
\affiliation{School of Natural Sciences, Technische Universit\"at München, 85748 Garching, Germany}

\author{Zi Yang Meng}
\email{zymeng@hku.hk}
\affiliation{Department of Physics and HK Institute of Quantum Science \& Technology, The University of Hong Kong, Pokfulam Road, Hong Kong SAR, China}

\begin{abstract}
The fascinating quantum many-body states in twisted bilayber graphene (TBG) at magic angle, due to the interplay of Coulomb interactions and the quantum metrics of flat bands, have been well understood both experimentally and theoretically. However, the phase diagram and excitations as functions of twist angle and permittivity 
are still largely unknown. Here, via a newly developed momentum-space continuous-field quantum Monte Carlo method fully taking into account long-ranged Coulomb interactions and flat bands' quantum metrics  with system sizes that were not accessible before, we show that charge-neutral TBG realizes an angle-tuned quantum phase transition from a gapped Kramers intervalley coherence (KIVC) state to a Dirac semimetal with critical angles around 1.2$\degree$. 
In single-particle spectra we demonstrate the evolution of a minimum gap at $\Gamma$ at the magic angle and 
towards touching points at Brillouin zone corners as the angle increases. 
The free energy and KIVC order parameter show that the transition belongs to fermionic Gross-Neveu criticality, and is robust upon varying the  permittivity or the interlayer hopping. 
\end{abstract}
\date{\today}
\maketitle

\section{Introduction}
Twisted bilayer graphene (TBG) has emerged as a tunable platform for studying the behavior of correlated electrons with quantum metrics. Coulomb interactions can dominate in TBG when narrow bands arise at some magic angles due to layer-hybridization and twist-induced band-energy shifts \cite{PhysRevLett.117.116804,PhysRevLett.108.076601,PhysRevLett.106.126802,SCHMIDT2010699,PhysRevB.81.165105,tramblyLocalization2010,bistritzerMoire2011,Santos2012,Datta2023Aug}. The narrow bands even become exactly flat and topologically non-trivial at 1.08$\degree$, the first magic angle (MA) in the chiral, particle-hole symmetric limit which arises for small angles without interlayer tunneling between equal-sublattice sites \cite{PhysRevLett.122.106405,bernevig2020tbg3,Sheffer_PRX_2023}. 
At the MA, a novel phase diagram as a function of filling and temperature was revealed in experiments including correlated insulators, superconductivity, orbital magnetism, and strange metallicity \cite{caoCorrelated2018,caoUnconventional2018,Lu2019,Wu2020,PhysRevLett.124.076801,stepanovCompeting2021,jaoui2022quantum,chen2024strong}. 

Further parameters such as strain, pressure, or magnetic field can also play important roles in TBG's non-trivial phase structure~\cite{yankowitz2019tuning,nuckolls2023quantum,PhysRevB.100.125104,PhysRevLett.127.027601,Kwan2021Kekule,wagnerGlobal2022,huangIntrinsic2022}. However, the twist angle, $\Theta$, as the perhaps most impactful tuning parameter is less explored. While first insights on the angle dependence are obtained from the comparison of different experiments or different domains in scanning tunneling measurements, recent developments made it possible to systematically tune the angle \emph{in-situ}, e.g., via a quantum twisting microscope~\cite{QTMShahal}, targeted bending~\cite{kapfer2023}, or manipulation of top rotators with an atomic-force-microscopy tip~\cite{ribeiro2018} or polymer handles~\cite{yang2020}. This raises the question of how the intricate interplay of the strong interaction, band geometry, and topology develops as a function of the twist angle and how angle-tuning can be exploited to optimize and explore phenomena beyond the currently known phase diagram. Theoretically, it is suggested that due to the effective change of interaction strength, angle-tuned TBG might exhibit the first realization of a quantum critical point (QCP) between the Kramer intervalley coherence (KIVC) insulator and a Dirac semimetal (DSM) \cite{partheniosTwisted2023,biedermannTwist2024} -- a condensed matter incarnation of the chiral phase transition of the Gross-Neveu (GN) model of quantum chromo dynamics~\cite{Semenoff_2012,ZINNJUSTIN1991105}, which has also recently been discussed in the context of moir\'e systems ~\cite{hawashinGross2025,tolosaRelativistic2025}.


Motivated by this, we here investigate the interacting band structure and phases of charge-neutral TBG in a range of twist angles via a newly developed unbiased continuous-field quantum Monte Carlo (CFMC) method where a continuum model of TBG with realistic parameters~\cite{tramblyLocalization2010,bistritzerMoire2011} is adopted. Better than the previously momentum-space quantum Monte Carlo (QMC) method for TBG developed by some of us~\cite{zhangMomentum2021,hofmannFermionic2022,zhangPolynomial2023,huangEvolution2024,panDynamical2022,zhangSuperconductivity2021,panThermodynamic2023}, here we employ continuous auxiliary fields and a global update scheme, inspired by the hybrid Monte Carlo method~\cite{beylRevisiting2018}, such that the computational complexity is greatly reduced by a factor of the system size and much higher momentum resolutions can be accessed. With such methodological innovation, we find a continuous phase transition from the correlated insulator close to the MA to a DSM as a function of the twist angle. 

\begin{figure}[!h]
\centering
\includegraphics[width=\linewidth]{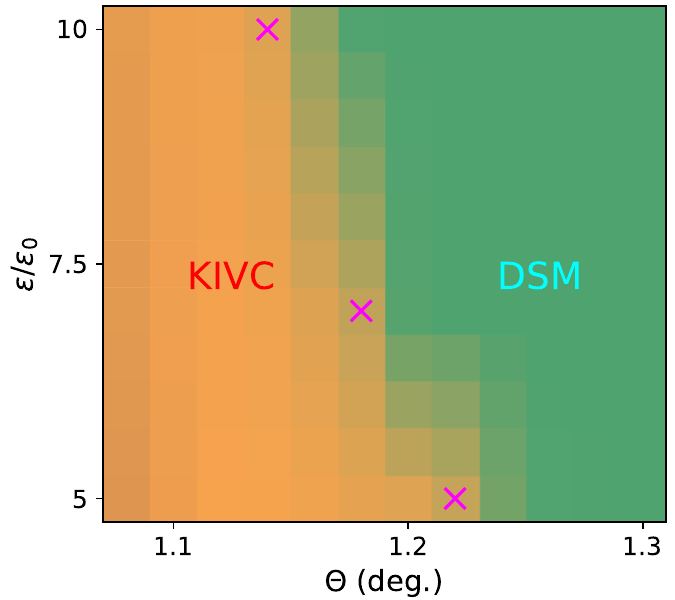}
\caption{\textbf{Phase diagram of pristine charge-neutral TBG versus twist angle $\Theta$ and permittivity $\varepsilon/\varepsilon_0$.} Here $u_0 = 60$ meV and $L = $ 15. The orange colored region is the KIVC ordered phase and the green colored region is the DSM phase. The color 
shading is based on our QMC data of the order parameters. The magenta crosses are the GN-QCP  determined from our QMC simulations (as shown in  \ref{fig:fig4} and consistent with Figs.~\ref{fig:fig2}, \ref{fig:fig3}). Close to the MA, grey admixture denotes the competing VP order, which is secondary to the KIVC order and is quickly suppressed by 
twist angle or system size. }
\label{fig:fig1}
\end{figure}

We show that in unstrained charge-neutral TBG, gaps due to KIVC \cite{poOrigin2018,liaoCorrelation2021,bultinck2020ground} decrease when increasing the twist angle from MA, and those of $\bK_{1,2}$ close continuously at $\Theta_{\c}=1.20(1)^\circ$ for the interlayer hopping parameters $u_0/u_1=0.545$ and the permittivity $\varepsilon/\varepsilon_0=7$, and the Dirac spectra are restored. Further increasing the twist angle leads to a DSM. Such a phase transition spontaneously breaks the chiral symmetry of Dirac fermions, falling into the chiral XY GN universality classes \cite{PhysRevLett.97.146401,MOSHE200369,ZINNJUSTIN1991105,ROSENSTEIN1993381}. Interestingly, as shown in the phase diagram of Fig.~\ref{fig:fig1}, we find the GN transition is robust upon tuning the permittivity $\varepsilon/\varepsilon_0$. Of course, the transition point will be shifted to larger $\Theta$ values as $\varepsilon$ decreases, i.e., the Coulomb interaction is enhanced. We also find that the transition is robust with different interlayer hopping parameters $u_0/u_1$ (shown in the Supplemental Material (SM)~\cite{suppl}). Moreover, close to the MA, the valley polarization (VP) order is also strong as a subleading instability, but when the twist angle is tuned away from MA, we find the twist angle quickly suppresses the VP order and helps to select the KIVC as the intrinsic critical mode for the Gross-Neveu transition. Our numerical observation can be well fitted into a field theoretical description, based on an effective Dirac fermion Lagrangian of the charge-neutral TBG and of Yukawa-Gross-Neveu type \cite{PhysRevD.10.3235} with emergent Lorentz invariance \cite{PhysRevB.79.085116,PhysRevB.80.081405,PhysRevB.80.075432,royLorentz}, the characteristic critical exponents are calculated from four-loop perturbation \cite{PhysRevD.96.096010,ihrigCritical2018} with $N_f=16$ fermion flavours from spin, valley, mini-valley, and sublattice. The obtained critical exponents make the correlations of KIVC order parameter with different system sizes well collapsed.  

\begin{figure*}[!ht]
\centering
\includegraphics[width=0.8\linewidth]{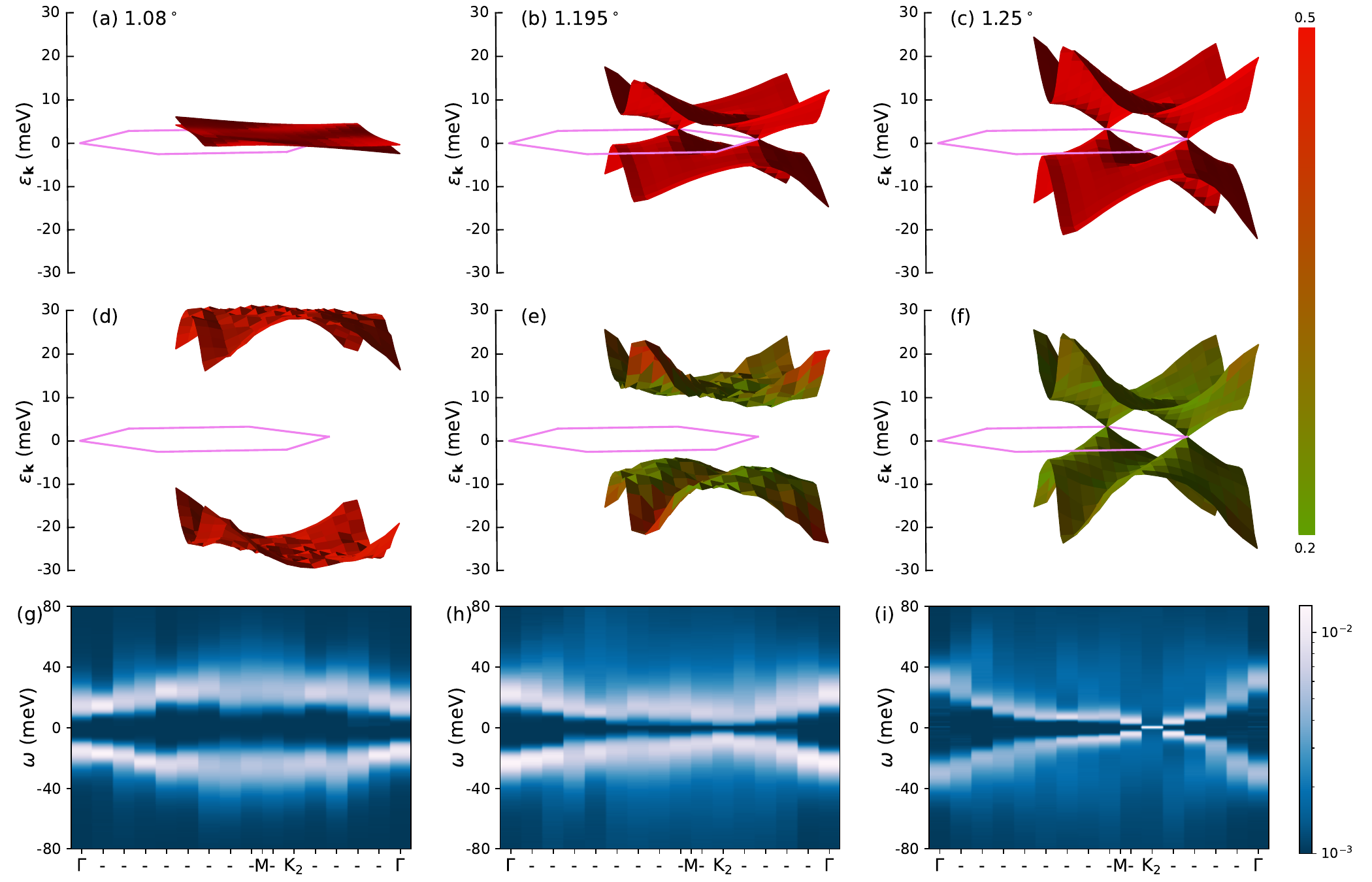}
\caption{\textbf{Single-particle excitation spectra as a function of the twist angle.} Here $\varepsilon = 7\varepsilon_0$ and $u_0 = 60$ meV. The violet hexagon is the mBZ with the momentum mesh of $15\times 15$ in each panel, and three columns are of $1.08\degree$, $1.195\degree$, and $1.25\degree$, respectively. (a)-(c) are the single-particle dispersion of the two low-energy bands from the continuum model without Coulomb interaction. At MA the two bands are almost flat, and the Dirac cones appear at $\bK_{1,2}$ of the mBZ. (d)-(f) are the CFMC obtained single-particle excitations with Coulomb interactions. They are obtained by a 2-mode approximation to the spectral function:  $-\frac{1}{\pi} \text{Im} \text{Tr} G^{\text{ret}}(\bk, \omega) \propto \sum_{m=1}^{2} a_m \delta(\epsilon_{m}(\bk) -  \omega)$, where $\epsilon_{m}(\bk)$ is the dispersion and $a_m$ the corresponding spectral weight denoted by the color coding. Panel (d) is at MA, where the system is gapped in the KIVC state and the minima of the single-particle gaps are at the $\Gamma$ point. CFMC obtained gaps at MA are consistent with those from the exact solution~\cite{zhangMomentum2021,bernevig2020tbg5} as also shown in Supplemental Material (SM)~\cite{suppl}. Panel (e) is at $\Theta=1.195\degree$, where the gaps at $\bK_{1,2}$ is about to close and the system is about to enter the Dirac semimetal phase through the GN-QCP. Panel (f) shows the dispersion at $1.25\degree$, where the Dirac dispersion is similar to that of the non-interacting counterpart in Panel (c), but with momentum-dependent weight distribution due to the Coulomb interaction. (g)-(i) Color mapping of the $ A(\bk,\omega)$ along a high-symmetric line, $\Gamma\rightarrow\mathrm{M}\rightarrow\bK_2\rightarrow\Gamma$.
} 
\label{fig:fig2}
\end{figure*}

\section{Results}

Our main results are summarized in the phase diagram of Fig.~\ref{fig:fig1}. As a function of the twist angle $\Theta$ and permittivity, $\varepsilon/\varepsilon_0$, the symmetry-breaking phase of KIVC and its GN-QCP to the DSM are identified. It is also noted that close to the MA, the secondary VP phase is present and it will be suppressed by the twist angle. We identify VP as secondary because its short-range correlation decreases with system size. Here in the main text we show the properties of the pristine TBG with the interlayer hoppings, $u_0 = 60$ meV and $u_1 = 110$ meV ($u_0/u_1=0.545$). Similar properties with different $u_0/u_1$ are shown in the SM~\cite{suppl}. Note that we fix $u_1 = $ 110 meV in this study. In the following, the model, the continuous-field QMC method, excitation spectra, free energy, and order parameter correlations, as well as their field-theoretical understanding and their experimental implications will be explained.

\subsection{Model and continuous-field Monte Carlo}
The kinetic part of our Hamiltonian comes from the continuum model of Bistritzer-MacDonald (BM)~\cite{bistritzerMoire2011} (see Methods for details). At MA of $\Theta=1.08\degree$, there emerges two low-energy bands, while the remote bands locates far away around 60 meV, with $u_0 = 60$ meV. The detailed dispersions of the continuum model are shown in Fig.~\ref{fig:fig6} in the Methods section. In the main text, we focus on the results of $u_0 = 60$ meV, and leave the outcomes of $u_0 = $ \ 30 and 90 meV in SM~\cite{suppl}, where the GN-QCPs between KIVC and DSM are similar. Note that for the non-interacting continuum model, Dirac cones are always present at the corners of the moir\'e Brillouin zone (mBZ) with momenta $\bK_{1,2}$. Enlarging the twist angle $\Theta$ 
magnifies the scale of the Dirac cones and broadens the two low-energy bands, but the remote bands are always separated by a sizable gap in the scale of 60 meV, as shown in Fig.~\ref{fig:fig6}.

In the 2D setting of TBG, the long-ranged single-gated (screened) Coulomb interaction for a momentum exchange $\bQ\neq0$ is given as $V(\bQ)/(4\Omega) =\frac{e^{2}}{16\Omega \pi \varepsilon} \int d^{2} \br\left(\frac{1}{\br}-\frac{1}{\sqrt{\br^{2}+d^{2}}}\right) \mathrm{e}^{\i \bQ \cdot \br}=e^2\left(1-\mathrm{e}^{-|\bQ| d}\right)/\left(8\Omega \varepsilon|\bQ|\right)$, with the permittivity $\varepsilon/\varepsilon_{0} = $5, 7, and 10, and $d/2=20$ nm the typical distance between TBG and a bottom gate~\cite{liuNematic2021}. Different $\varepsilon$ changes the overall strength of the Coulomb interaction, as shown in Fig.~\ref{fig:fig1} the phase diagram spanned by the axes of $\varepsilon/\varepsilon_0$ and $\Theta$. $\Omega$ is the total area of the real-space moir\'e superlattice. Since the scale of the Coulomb interaction is less than 6 meV as $|\bQ|\rightarrow0$~\cite{panThermodynamic2023,bernevig2020tbg5}, as also shown in the SM~\cite{suppl} in Fig.~\ref{fig:V(0)}, 
it will not mix the remote bands with the two low-energy bands in the cases considered in this study, and we can project the Hamiltonian from plane-wave basis to band basis considering only the two low-energy bands. Then, the Hamiltonian reads 
\begin{equation}
\begin{aligned}
H=&\sum_{s,\eta,\bk,m}\epsilon^{s,\eta}_{\bk,m}c^\dagger_{s,\eta,\bk,m} c_{s,\eta,\bk,m}+\sum_\bQ\frac{1}{4\Omega}V(\bQ)\left(A^2_\bQ-B^2_\bQ\right)\\
=&H_0+H_\mathrm{I},
\end{aligned}
\label{eq:eq1}
\end{equation}
where $\epsilon^{s,\eta}_{\bk,m}$ is the kinetic energy for spin $s = \uparrow, \downarrow$, valley $\eta = \pm$, momentum $\bk$, and band $m$. The interaction term $A_\bQ=\delta_{\rho_{-\bQ}}+\delta_{\rho_\bQ}$ and $B_\bQ=\delta_{\rho_{-\bQ}}-\delta_{\rho_\bQ}$ with the density operator 
\begin{equation}
\begin{aligned}
\delta_{\rho_\bQ}=&\sum_{s,\eta,k,m,n}\lambda^{s,\eta}_{m,n}\left(\bk,\bQ\right)\\
&\left(c^\dagger_{s,\eta,\bk,m}c_{s,\eta,\bk+\bQ,n}-\frac{4+\nu}{8}\delta_{q,0}\delta_{m,n}\right),
\end{aligned}
\label{eq:eq2}
\end{equation}
and the filling factor $\nu$. The form factor $\lambda^{s,\eta}_{m,n}$ comes from the projection from plane-wave basis to band basis~\cite{huangEvolution2024,zhangMomentum2021} (see Methods for detailed derivation of Eq.~\eqref{eq:eq1}). As shown in our previous work~\cite{panThermodynamic2023}, $V(\bQ)/(4\Omega)$ decays to almost 0 at the distance of $|\mathbf{G}_{1,2}|$, the module of lattice constants of mBZ, where a cutoff can be made. In the CFMC simulation, we discretize the mBZ into an $N=L\times L$ momentum grid for $\bk$.

A momentum-space QMC scheme for Eq.~\eqref{eq:eq1} has been developed in Refs.~\cite{zhangMomentum2021,hofmannFermionic2022}. However, due to the long-range nature of the Coulomb interactions and the form factor that generates all-to-all momentum transfers, the QMC computation for TBG suffers from a very high computational complexity of at least $O(\beta N^4)$, where $\beta=1/T$ is the inverse temperature (see the detailed analysis in the Methods section). So, previous QMC works on TBG are all confined in relatively small systems sizes (usually up to $L=9$) and less scanning of the phase diagram in the parameter space~\cite{zhangMomentum2021,hofmannFermionic2022,zhangPolynomial2023,huangEvolution2024,panDynamical2022,zhangSuperconductivity2021,panThermodynamic2023}. However, to be able to discuss the spectral properties and universality originating from the Dirac fermions, one needs to be able to simulate larger systems, and therefore the reduction of the computational complexity becomes crucial. As pointed out in Ref.~\cite{ippolitiHalf2018}, the scaling can be improved by considering continuous auxiliary fields and global updating schemes such as Hybrid Monte Carlo~\cite{beylRevisiting2018}. This is precisely what we have achieved in this work, as detailed in the Methods section on the CFMC algorithm with both Gaussian update scheme and Hamiltonian update scheme, which successfully reduces the computational complexity to $O(\beta N^3)$ and allows us to access a large system with $L=15$ at low temperature. Such large-scale QMC simulations are carried out at each set of parameters of $\{\varepsilon, u_0\}$ while $\Theta$ is continuously tuned. The obtained single-particle spectrum, thermodynamic free energy, and various order parameters confirm the chiral XY GN quantum criticality in TBG.
 
\begin{figure}[!h]
\centering
\includegraphics[width=\linewidth]{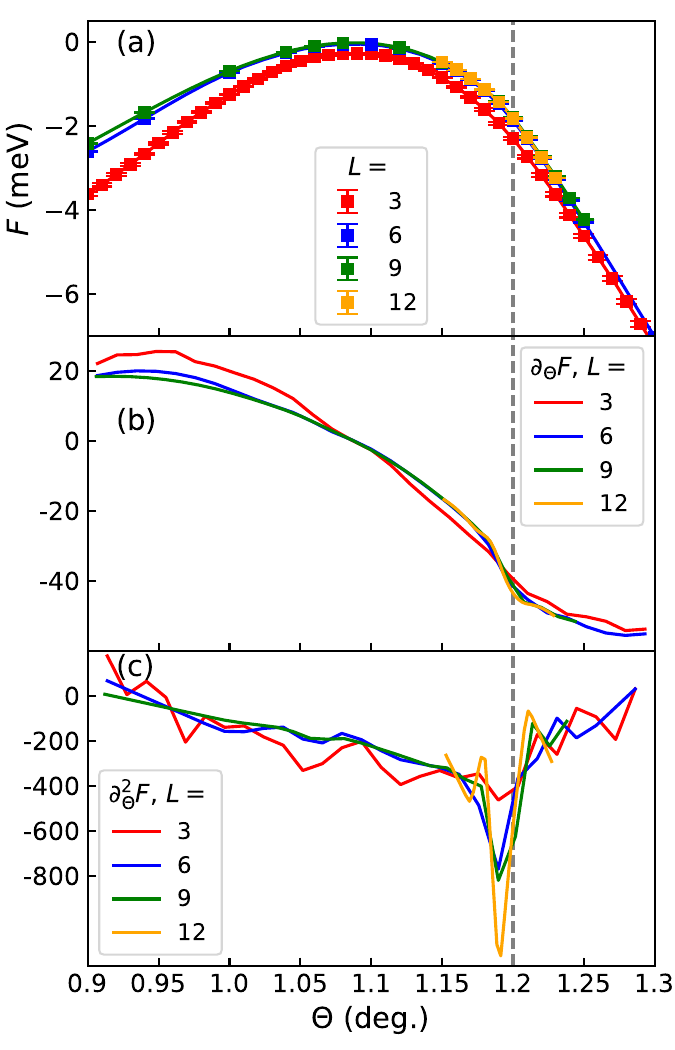}
\caption{\textbf{Indication of the GN-QCP from the free energy.} Free energies of $L=$ 3, 6, 9, and 12 systems (with $\beta \propto L$) (a) and their first- (b) and second-order (c) derivatives for $\varepsilon = 7\varepsilon$ and $u_0 = $ 60 meV. The second-order derivatives exhibit a divergence at $\Theta_\c\sim1.2\degree$ as a function of the system size, signaling a second-order phase transition between the KIVC insulator and the DSM. Note that the lines in (a) are spline fits of the data points (whose errorbars are smaller than the symbol size), from which the derivatives in (b) and (c) are performed.}
\label{fig:fig3}
\end{figure}

\subsection{Single-Particle Spectrum, Free Energy and Angle-Tuned GN criticality}

In Fig.~\ref{fig:fig2} we show the single-particle spectra without ((a)-(c)) and with ((d)-(i)) Coulomb interactions, where $\Theta$ is tuned from 1.08$\degree$ to 1.25$\degree$, and we focus on the parameters of $\{\varepsilon=7\varepsilon_0, u_0=60$ meV\}. As will be discussed later, for other parameter sets of $\{\varepsilon, u_0\}$, order parameter analyses reveal similar behavior. The non-interacting dispersion ((a)-(c)) is from the BM model (see Methods) and the interacting single-particle spectra ((d)-(i)) are obtained from CFMC simulations. These spectra are obtained by a 2-mode approximation to the spectral function (we trace out the 2 $\times$ 2 matrix of dynamical Green's function at each momentum point in the band basis):  $-\frac{1}{\pi} \text{Im} \text{Tr} G^{\text{ret}}(\bk, \omega) \propto \sum_{m=1}^{2} a_m \delta(\epsilon_{m}(\bk) -  \omega)$, where $G^{\text{ret}}(\bk, \omega)$ is the single-particle Green's function from the CFMC simulation, $\epsilon_{m}(\bk)$ the dispersion, and $a_m$ the corresponding spectral weight denoted by the color coding in Fig.~\ref{fig:fig2}. 

At the MA 1.08$\degree$, 
the non-interacting bandwidth of TBG is on the scale of 1 meV, as shown in Fig.~\ref{fig:fig2} (a). Thus, the system is dominated by Coulomb interactions on a scale of 5 meV for $\varepsilon = 7\varepsilon_0$. As a result, charge-neutral TBG manifests a correlated insulator of KIVC as its ground state~\cite{caoCorrelated2018,caoTunable2020,polshynLarge2019,liuTuning2021,xie2019spectroscopic,choi2019electronic,nuckolls2020strongly,
saito2021hofstadter,das2021symmetry,wu2021chern,lianTwisted2021tbg4,xieTwisted2021,bultinck2020ground,potaszExact2021,
wilhelmInterplay2021,xieNature2020,YiZhang2020,liuTheories2021,hejaziHybrid2021,xiePhase2023,linSymmetry2020,kwanDomain2021,dattaHeavy2023,
kangNon-Abelian2020,soejimaEfficient2020,zhangMomentum2021,panDynamical2022,hofmannFermionic2022,panThermodynamic2023,zhangPolynomial2023,
saito2020independent,saito2021hofstadter,luSuperconductors2019,xie2019spectroscopic,choi2019electronic,stepanov2020untying,polshynLarge2019,stepanov2020untying,
liuNematic2021,Rai2024Sep}, as shown in Fig.~\ref{fig:fig2} (d), where the CFMC dispersion is consistent with that from exact solution of the purely interacting Hamiltonian, $H_\mathrm{I}$, and previous momentum-space QMC results~\cite{zhangMomentum2021,huangEvolution2024,bernevig2020tbg5}. When $\Theta$ is tuned further away from 1.08$\degree$, the non-interacting bandwidth evolves to $\sim$15 meV at 1.195 $\degree$ and $\sim$23 meV at 1.25 $\degree$, as shown in Fig.~\ref{fig:fig2} (b) and (c) respectively, so that the kinetics overwhelm the Coulomb interactions. As a result we find a strong qualitative change of the full single-particle QMC spectra. Already slightly away from MA, they cannot be approximated anymore by a purely interacting system, see Fig.~\ref{fig:exact} in SM \cite{suppl}. The minimal excitation energy moves from the mBZ origin $\Gamma$ to mBZ corners $\bK_{1,2}$ as a function of twist angle. Gaps are 
decreasing at $\bK_{1,2}$ at 1.195 $\degree$ as in Fig.~\ref{fig:fig2} (e), and totally closed at 1.25 $\degree$ with Dirac band touching as in Fig.~\ref{fig:fig2} (f) and (i), where even in the presence of interactions, the system is a DSM. However, spectral weight redistributes to higher energy. The DSM is known to be stable against perturbative interactions due to the vanishing density of states at Fermi level. Therefore, a quantum phase transition is expected between the correlated insulator and the DSM, and if the transition turns out to be continuous, it 
corresponds to a GN-QCP.

Spectral functions $A(\bk,\omega)$ are also computed 
using the stochastic analytic continuation (SAC)~\cite{Sandvik1998Stochastic,beachIdentifying2004}, the computational scheme of which is explained in detail in Methods section. Such a QMC+SAC scheme has been successfully applied in previous momentum-space QMC simulations to reveal the ground state and finite-temperature single-particle and collective excitation spectra of the TBG systems~\cite{panDynamical2022,zhangSuperconductivity2021,huangEvolution2024,panThermodynamic2023}. Our obtained spectra for the $15\times 15$ system are shown in Fig.~\ref{fig:fig2} (g), (h) and (i), where we have chosen a high-symmetry path inside the mBZ that scans through the $\bK_2$ point to highlight the different natures of the states across the transition. The peak positions of $A(\bk,\omega)$ are in full consistency with those from the 2-mode approximation in Fig.~\ref{fig:fig2} (d), (e) and (f).  Fig.~\ref{fig:fig2} (g) is inside the KIVC state at MA and the spectrum shows clear upper and lower Hubbard bands for the insulating ground state. Fig.~\ref{fig:fig2} (h) is close to the critical angle $\Theta_\c$ and the insulating gap is closing at the Dirac point $\bK_2$. Fig.~\ref{fig:fig2} (i) is at $1.25\degree$ thus inside the DSM phase, and the linear dispersion originating from the Dirac point manifests itself. 

\begin{figure*}[!ht]
\centering
\includegraphics[width=\linewidth]{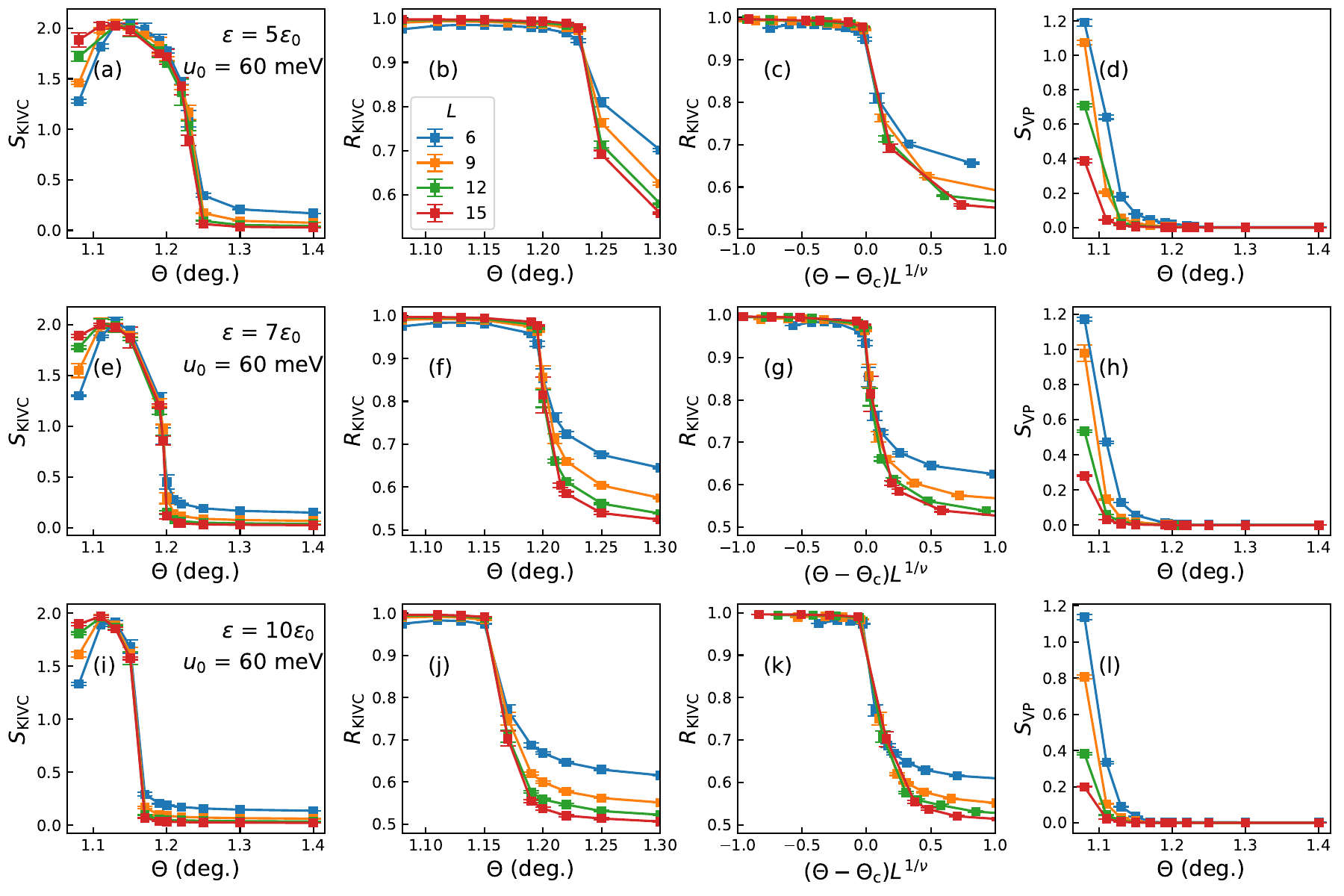}
\caption{ \textbf{The GN-QCP analysis from the KIVC order with different permittivities}. $u_0 = 60$ meV is fixed while $\varepsilon/\varepsilon_0$ is altered. For permittivities of 5, 7 and 10, respectively, each row contains the structure factor $S$ of the KIVC order, its correlation ratio, finite-size collapse, and $S$ of the VP order, on the lattices with $L=6, 9, 12$, and 15 as a function of $\Theta$. We set $\beta \propto L$ in these CFMC simulations. From $1.08\degree$ to $1.4\degree$, the KIVC order continuously vanishes (although there are strong competition from VP when $\Theta<1.15\degree$) as the systems evolve from KIVC insulator to DSM. Correlation ratio gives the critical angle, $\Theta_{\c}= 1.233\degree, 1.197\degree, \mathrm{and} \ 1.156\degree$, respectively for $\varepsilon/\varepsilon_0 = $ 5, 7, and 10. The finite-size data collapses are consistent to the critical exponent, $\nu = 1.13$ of the $N_f = 16$ GN-QCP.}
\label{fig:fig4}
\end{figure*}

To locate the exact position and to explore the nature of the transition, we compute the free energy $F(\Theta)$ of the system as a function of $\Theta$. The results are shown in Fig.~\ref{fig:fig3}. Free energy, as one of the exponential observables~\footnote{This is because to compute the free energy, one need to evaluate $\langle \e^{-F} \rangle$ instead of the $F$ in the path-integral QMC, and that is an exponentially small number, hence the name of exponential observable. Incremental method is one way to carry out its calculation gradually.}, can be computed at finite temperature with the incremental method developed by some of us~\cite{zhangIntegral2024}. Details of the algorithm are given in the Methods section. Fig.~\ref{fig:fig3} (a) shows that $F(\Theta)$ is a smooth function for $\Theta \in [0.9\degree, 1.3\degree]$ (the inverse temperature is set to be $\beta \propto L$ to access the ground state properties as the system size increases). The lines in Fig.~\ref{fig:fig3} (a) are spline fits of the data points. Fig.~\ref{fig:fig3} (b) shows that the first-order derivative of the spline-fitted $F(\Theta)$ versus $\Theta$, $\partial F(\Theta)/\partial\Theta$, is also a smooth function, and only the second-order derivative, ${\partial^2 F(\Theta)}/{\partial \Theta^2}$, exhibits the signature of a divergence at $\Theta \sim 1.2\degree$ in Fig.~\ref{fig:fig3} (c). These results indicate that the phase transition is continuous and the GN-QCP is a likely scenario in the vicinity of $1.2\degree$. 

The correlated insulating state of KIVC breaks the U(1) symmetry of the charge conservation between the valley degrees of freedom~\cite{liaoCorrelated2021,liaoCorrelation2021,bultinck2020ground,hofmannFermionic2022}. Considering the degrees of freedom of two spins ($s$), two valleys ($\eta$), two mini-valleys ($\tilde\mu$), and two 
sublattices ($\sigma$), the Dirac semimetal at the charge neutrality point will have $N_f=16$ component Dirac fermions, i.e., four flavors of four-component Dirac fermions. Since the semimetal will recover the U(1) (XY) symmetry broken by the KIVC insulator, an $N_f=16$ chiral XY GN-QCP is expected.
The effective Lagrangian describing this  universality class is given by
\begin{align}
    \mathcal{L}&=\int\!\!\! \frac{
    d\omega d^2k}{(2\pi)^3}\,\psi^{\dagger}(-i\omega+ 
    \eta_z \sigma_x k_x + \sigma_y k_y)\psi \nonumber \\ 
    &+ \int\!\!\! d^3x \, \left[\frac{1}{2}\phi(\partial^2_{x}+\partial^2_{y})\phi 
    +\frac{1}{2}m^2\phi^2+\frac{1}{4}\lambda\phi^4+g\phi^a\psi^{\dagger} M_a\psi \right]
    \label{eqn:GNYLag}
\end{align}
where $\psi$ and $\phi$ denote the fermionic and bosonic (order) parameter fields and $a$  = 1, 2 the two real components 
of $\phi$. The 16-component low-energy fermion spinor $\psi_\bk=(c_{s,\eta,\bK_1+\bk, \sigma},c_{s,\eta,\bK_2+\bk,\sigma})^\mathrm{T}$ is obtained by expanding fermion modes around the mini-valley $\bK_{1,2}$ points. Note that due to the non-trivial form of the BM Hamiltonian, we use a generalized sublattice $\sigma$ that describes the effective low-energy Hamiltonian obtained from this expansion. In the limit where only one shell of reciprocal lattice vectors is considered, one can show analytically that the sublattice degree of freedom exactly corresponds to the microscopic graphene sublattices \cite{bistritzerMoire2011,TBG1}. 
In our case, for the chiral XY phase transition, $M_1=\eta_x\sigma_x\tilde\mu_y$ and $M_2=\eta_y\sigma_x\tilde\mu_y$ where $\tilde\mu$. In the limit where the inclusion of the inter-layer tunneling is perturbative, at the $\mathbf{K}_{1,2}$ points of the moir\'e BZ, the mini-valley degree of freedom can be identified with the layer. 

The consistency to a chiral XY GN-QCP is verified by the correlation of KIVC order parameter computed in the CFMC simulation. The order parameter of KIVC is defined from the plane-wave basis with proper projections to the band basis, as
\begin{equation}
\begin{aligned}
O_\mathrm{KIVC}(\bq)\equiv&\sum_{s,\bk}\vd^\dagger_{s,\bk+\bq}\eta_x\mu_y\sigma_x\vd_{s,\bk}\\
=&\sum_{s,\eta,\bk,m,n}c_{s,\eta,\bk+\bq,m}^\dagger\langle u^{s,\eta}_{\bk+\bq,m}|\mu_y\sigma_x|u^{s,-\eta}_{\bk,n}\rangle c_{s,-\eta,\bk,n},
\end{aligned}
\end{equation}
with $\mu_y$ acting on layers, $\sigma_x$ acting on sublattices, and $|u^{s,\eta}_{\bk,m}\rangle$ the eigenvector relating to the low-energy band $m$. In a similar way, the order parameter of VP, as a competing order around MA~\cite{panDynamical2022}, is defined as
\begin{equation}
\begin{aligned}
O_\mathrm{VP}(\bq)\equiv&\sum_{s,\bk}\vd^\dagger_{s,\bk+\bq}\eta_z\mu_0\sigma_0\vd_{s,\bk}\\
=&-\sum_{s,\eta,\bk,m,n}\eta c_{s,\eta,\bk+\bq,m}^\dagger\langle u^{s,\eta}_{\bk+\bq,m}|u^{s,\eta}_{\bk,n}\rangle c_{s,\eta,\bk,n}.
\end{aligned}
\end{equation}
The conversion from the plane-wave basis to band basis, for the measurements of KIVC and VP orders, especially in terms of the gauge choice introduced by the projection, are discussed in details in the Methods section~\cite{hofmannFermionic2022,bultinck2020ground,liaoCorrelation2021,panDynamical2022}.

The correlation function of the KIVC and VP orders, $S(\bq)=\langle {O(\bq)}^\dagger O(\bq)\rangle/N^2$, at its ordered wavevector $\bq=\Gamma$ as a function of $\Theta$, is shown in the first and fourth columns of Fig.~\ref{fig:fig4} for $\varepsilon/\varepsilon_0 = $ 5, 7, and 10, respectively in each row. The orders at MA have been intensively studied~\cite{liaoCorrelated2021,bultinck2020ground,liaoCorrelation2021,hofmannFermionic2022,panDynamical2022}, and it is known that due to the presence of kinetic energy, the KIVC order is the leading instability compared over the VP order~\cite{panDynamical2022,hofmannFermionic2022}. From our data in Fig.~\ref{fig:fig4}, it is clear that KIVC order is strong at $\Theta<1.2\degree$. At MA, the VP order is also pronounced although it is only secondary -- the strength of $S_\mathrm{VP}$ decreases with system size, while the strength of $S_\mathrm{KIVC}$ saturates as $L$ increases. 
It is this competition that weakens the KIVC and consequently make it most pronounced not at MA but at angles above it, $\sim1.13\degree$ for the three different $\varepsilon/\varepsilon_0$. At such angles, the VP order already vanishes, so that the twist angle here plays the role of lifting the competing phase and singling out the KIVC order as the critical mode of the GN transition. As the twist angle increases further, the KIVC order vanishes around $1.2\degree$ in a steep but continuous way, which is consistent with the free-energy behaviour within our resolution for angles. Similar behavior at different interlayer hopping parameters $u_0=30$ and 90 meV, are shown in the SM~\cite{suppl}.

To investigate the critical behavior of this transition, we calculate the critical exponents of the $N_f=16$ chiral XY GN-QCP. More specifically, the inverse correlation length exponent $\nu^{-1}$ and the anomalous dimension of the KIVC order parameter field $\eta$ are obtained by expanding around the upper critical (space-time) dimension $D=4-\epsilon$. We use the results of a four-loop $\epsilon-$ expansion \cite{PhysRevD.96.096010} and average over different Pad\'e approximants for the dimensional dependence (details of the calculation are given in the Methods). A hyperscaling relation is employed to determine the scaling exponent of the order parameter $\beta=\nu{(D-2+\eta)}/{2}$. We find $\nu=1.13$ and $\beta=1.09(5)$. Furthermore, within the four-loop $\epsilon$-expansion, we are able to calculate the correction-to-scaling exponent $\omega$, which corresponds to an $L^{-\omega}$ term in the scaling function of the KIVC order parameter and for $N_f=16$, $\omega=0.862(1)$ is obtained. For comparison, $\omega=0.788(8)$ for the same universality class in monolayer graphene, which corresponds to $N_f=8$. As such, we expect these corrections to be negligible due to the larger number of fermion flavours that are present in TBG.

Next, we calculate the correlation ratio of KIVC, $R_\mathrm{KIVC}=1-S(\bq)/S(\Gamma)$, where $S(\bq)$ is the average value of six $\bq \sim 1/L$ nearest to $\Gamma$, as shown in the second column of Fig.~\ref{fig:fig4}. The values of $R_\mathrm{KIVC}$ for different system sizes lead to finite-size crossings, with the obtained critical angles, $\Theta_{\c}=1.233\degree, 1.197\degree, \mathrm{and} \ 1.156\degree$, respectively, for $\varepsilon/\varepsilon_0=5, 7, \mathrm{and}\ 10$. In the third column of Fig.~\ref{fig:fig4}, the correlation ratios collapse well around corresponding critical angles with the GN exponent $\nu=1.13$ for all permittivities. The counterpart of Fig.~\ref{fig:fig4} with altering $u_0$ and fixing $\varepsilon/\varepsilon_0=7$ is shown in Fig.~\ref{fig:R1} in SM~\cite{suppl}, which also indicates $\nu=1.13$ with different critical angles $\Theta_{\c}=1.159\degree, 1.197\degree, \mathrm{and} \ 1.254\degree$, respectively for $u_0=30, 60, \mathrm{and}\ 90$ meV.


These results demonstrate the consistency with the chiral XY GN-QCP with the position and nature of the transition in TBG as indicated in the single-particle spectrum and free energy. We note that such consistency is only made possible by the newly developed CFMC method, which can access large system sizes (up to $L=15$). Otherwise, it is impossible to probe the nature of this GN-QCP with the discretized-field version of momentum-space QMC, in which the largest system size is $L=9$. In addition, in the SM~\cite{suppl}, we also show 
an attempted data collapse with the (2+1)D XY exponent $\nu=0.67$~\cite{chesterCarving2020,campostriniCritical2001}, 
which, as expected, turns out to be much less satisfactory. This reinforces the understanding that the transition we observe is non-thermal and involves the Dirac fermions, i.e., it is of GN-type.

\section{Discussion}
With the newly developed CFMC method, we succeeded in discovering an angle-tuned quantum critical point of Gross-Neveu type in pristine TBG at charge neutrality, which separates semi-metallic Dirac electrons from a gapped phase with the KIVC order. Our observation is robust in the sense that with different permittivities $\varepsilon/\varepsilon_0$ and the interlayer hopping $u_0$, we have observed the same critical behavior of a GN-QCP at different $\Theta_\c$. We believe such an interaction-driven transition is not possible to realize in graphene because the effective interaction is too weak. Therefore, we offer a new and exciting opportunity that if one can experimentally tune the angle from the magic one (1.08$\degree$) to 1.3$\degree$ at low temperature, a quantum critical point with novel critical behavior and an emergent Dirac semimetal can be seen in pristine TBG. 

The tuning of the twist angle alters the scale of the low-energy bands and their interactions. As $\Theta$ increases, the Coulomb interactions effectively become weaker in comparison with the kinetic energy, which has a strong effect on the single-particle excitations in TBG. When $\Theta < \Theta_\c$, single-particle excitations are well approximated by exact solutions of the purely interacting Hamiltonian, and with a minimum at $\Gamma$ at the magic angle $\Theta=1.08\degree$. However, the spectra change qualitatively as a function of the angle toward a Dirac dispersion with touching points at $\bK_{1,2}$ for $\Theta>\Theta_\c$. During the evolution, the spectral weights from high-energy modes towards the low-energy Dirac cones redistribute. 
As an interesting by-product, we notice that the critical region is small in terms of the control parameter $\Theta$, which is probably due to that the projected Coulomb repulsion generates many possible mass terms that compete. This is reminiscent of single layer graphene with unscreened Coulomb repulsion where both charge density wave and  antiferromagnetic orders are strong~\cite{Tang17} and similar behavior of a narrower window of the transition has been observed~\cite{hohenadlerPhase2014,PhysRevB.98.235129}. 

Moreover, we saw the interesting phenomenon that close to MA, the symmetry-breaking phases KIVC and VP are competing. In our settings, the KIVC is the leading order and the correlation of VP vanishes with increasing system size. As $\Theta$ increases, independent of the values of ${\varepsilon}/{\varepsilon_0}$ and ${u_0}/{u_1}$, the 
the competition is suppressed and 
KIVC is singled out as the long-range critical mode for the GN-QCP 
at charge neutrality. 
This observation implies that, although the TBG sample at MA is difficult to prepare, 
if one wants to pursue novel quantum 
phase transitions in TBG, one can purposely go away from the MA and the twist angle will help to identify the suitable critical mode. 

In terms of quantum many-body computation, it is well-known that the progress in numerical simulations of TBG is important, and the momentum-space QMC and its CFMC improvement implemented in this work are substantial developments. The usage of the CFMC algorithm can be even broader. In fact, the algorithms for TBG are similar to those used in the realm of Landau level regularization schemes for continuum field theories, let it be on the torus \cite{ippolitiHalf2018,wangPhase2021} or on the sphere \cite{Hofmann24,chenPhases2024}. Also, our CFMC method has potential applications in MoTe$_2$ and multilayer graphene  systems where the fractional quantum anomalous Hall effects have recently been observed~\cite{caiSignature2023,park2023_fqah,multilayer_graphene_fqah,luExtended2024,luThermodynamics2024}. In TBG and other 2D quantum moir\'e systems, the interplay of the long-range Coulomb interaction and the quantum metric of flat bands must be considered with equal footing, and our approach paves the way to achieve this with much larger system sizes. 

At the fundamental level, our proposal of the GN-QCP realization in TBG is more generally relevant to the study of interaction-induced phase transitions in Dirac semimetals. Close to the transition point, the emergent quantum critical behavior is universal. We demonstrated that the critical behaviour in TBG is consistent with the Gross-Neveu chiral XY universality class with emergent Lorentz invariance. The Gross-Neveu model describes the spontaneous mass generation of relativistic Dirac electrons, and can be taken as an effective low-energy model for the chiral phase transition of quantum chromo dynamics. In the condensed matter context, its quantum phase transition was studied extensively via different theoretical methods such as QMC~\cite{Chandrasekharan-Quanum-2013,He-Dynamical-2018,Liu-Designer-2020,Tabatabaei-Chiral-2022,Bonati-Chiral-2023,liaoValence2019,assaadPinning2013,wangQuantum2023,liaoDiracIII2022,ihrigCritical2018,PhysRevB.94.245102,PhysRevD.96.114502,Li_2015,PhysRevD.103.065018,Iliesiu-Bootstrapping-2018,PhysRevB.102.235105,doi:10.1142/S0217751X94000340,PhysRevD.97.105009,PhysRevD.96.096010,PhysRevLett.123.137602,PhysRevB.97.075129}, bootstrap~\cite{Erramilli-Gross-2023,Iliesiu-Bootstrapping-2018}, and renormalisation group~\cite{herbutInteraction2006,graceyFour2016,mihailaGross2017,ihrigCritical2018,PhysRevD.96.096010}, and critical exponents are beginning to converge. In this sense, it can be seen as a minimal model for fermionic quantum critical transitions, analogous to the O(N) model at thermal phase transitions. Thus, the experimental realization in a highly-tunable platform like TBG promises the possibility to simulate, on the one hand, a standard model for fermionic quantum criticality, and on the other hand, the spontaneous mass generation during chiral symmetry breaking in two dimensions.  


Experimentally, \emph{in-situ} tuning of the twist angle has recently become possible, e.g., via a quantum twisting microscope~\cite{QTMShahal}, or manipulation via microscopic handles~\cite{kapfer2023,ribeiro2018,yang2020}. Furthermore, information on the angle dependence of single-particle spectra can be obtained from quantum twisting microscope, nano angle-resolved photoemission spectroscopy (nano-ARPES)~\cite{chen2024strong}, and from different domains in scanning tunneling microscopy (STM). The presence of Dirac electrons and their velocity renormalization with twist angle can be measured via quantum oscillations. A spectral gap can be extracted from transport, nano-ARPES, STM and quantum capacitance experiments, which can also help to infer different symmetry breaking due to interactions. In this context, it is also important to note that strain has a strong effect on the spectrum and can drive the system back into a semimetallic phase. It will be interesting to study further how strain affects the angle-tuned quantum 
phase transition. 

Addition note: During the preparation of this work, Ref.~\cite{biedermannTwist2024} appeared on the arXiv, in which the angle-dependent quantum critical behavior is studied within unrestricted Hartree-Fock. Although the values of the critical angle are slightly different, due to different model parameters, the conclusions therein are in line with our findings.

\vspace{1cm}

\noindent{\textbf{\large Methods}}

\noindent{\textbf{Sec. I \ Continuum model of angle-twisted TBG.}} 
\begin{figure}[!h]
\centering
\includegraphics[width=\linewidth]{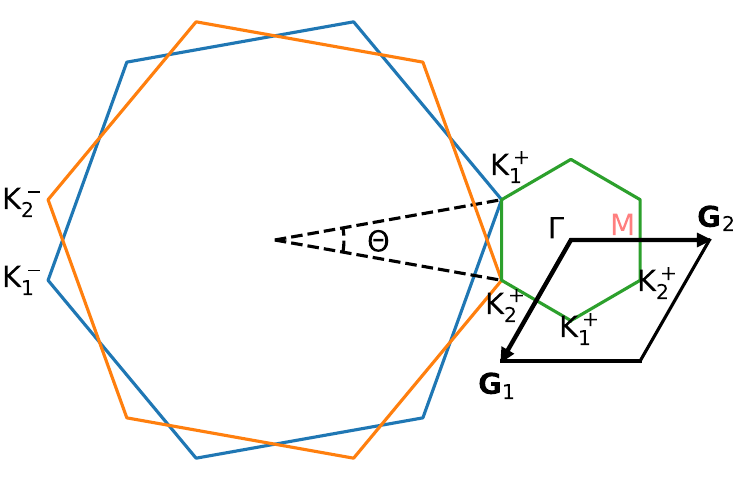}
\caption{\textbf{Momentum-space configuration of BZs and mBZs}. Twisting of graphene BZs as blue and orange hexagons leads to the mBZs as green hexagon or black rhombus with lattices of $\mathbf{G}_1$ and $\mathbf{G}_2$. $\Theta$ is the twist angle and $\bK^\eta_l$ is a Dirac point of valley $\eta$ and layer $l$. $\Gamma$ and M are two high-symmetric points.}
\label{fig:fig5}
\end{figure}

The schematic twist of graphene Brillouin zone and the consequent moir\'e Brillouin zone are drawn in Fig.~\ref{fig:fig5}. The length of the lattice vectors of mBZ is $|\mathbf{G}_{1}|, |\mathbf{G}_2|=8\pi\mathrm{sin}(\Theta/2)/(3a)$ with $a=1.42 \ \si{\angstrom}$ being the distance between the nearest carbon atoms. $\bK_1^\eta$ and $\bK_2^\eta$ are Dirac points of valley $\eta=\pm$ from the top (1) and bottom (2) layers, respectively. A momentum-space single-particle Hamiltonian, $H_0$, considering intralayer hopping and interlayer hopping, can be constructed in the plane-wave basis as
\begin{equation}
\begin{aligned}
H_0=&\sum_{s,\eta,\bk,\bG,X,\bG',X'}d^\dagger_{s,\eta,\bk,\bG,X}H^{s,\eta,\bk}_{\bG,X;\bG',X'}
d_{s,\eta,\bk,\bG',X'}\\
=&\sum_{s,\eta,\bk}\vd^\dagger_{s,\eta,\bk}H^{s,\eta,\bk}
\vd_{s,\eta,\bk}.
\end{aligned}
\label{eq:eq0}
\end{equation}
Here $d^\dagger_{s,\eta,\bk,\bG,X}$ is the creation operator for spin $s$, valley $\eta$, momentum $\bk$ in mBZ, momentum folding $\bG$ to account for the contributions from extended Brillouin zones, layer (1, 2) and sublattice (A, B), and $\vd^\dagger_{s,\eta,\bk}$ is the vector of $d^\dagger_{s,\eta,\bk,\bG,X}$. We use $X\in\{1\mathrm{A}, 1\mathrm{B}, 2\mathrm{A}, 2\mathrm{B}\}$ to denote the layer and sublattice degrees of freedom. So, $H^{s,\eta,\bk}$ is a matrix with the dimension of $N_\bG N_X$, which is actually the BM model~\cite{bistritzerMoire2011}, and with the sub-blocks as
\begin{equation}
	\begin{aligned}
		&H^{s,\eta,\bk}_{\bG,\bG'}\\
        &=\delta_{\bG,\bG'}\hbar\nu_\mathrm{F}
\left(\begin{array}{cc} 
			-(\bk+\bG-\bK_1^\eta) \cdot \pmb{\sigma}^\eta  &  0  \\
			0  & - (\bk+\bG-\bK_2^\eta) \cdot \pmb{\sigma}^\eta
		\end{array}\right) \\
		&+\left(\begin{array}{cc} 
			0  &  T^\eta_1  \\ 
			{T^\eta_2}^\dagger & 0
		\end{array}\right)
	\end{aligned}
	\label{eq:BM}
\end{equation}
with
\begin{equation}
\begin{aligned}
T^\eta_l=&\left(\begin{array}{cc}
u_0 & u_1\\
u_1 & u_0\end{array}\right)\delta_{\bG,\bG'}
+\left(\begin{array}{cc}
u_0 & u_1\mathrm{e}^{-i\frac{2\pi}{3}\eta}\\
u_1\mathrm{e}^{i\frac{2\pi}{3}\eta} & u_0\end{array}\right)\delta_{\bG,\bG'+(-1)^l\eta\G_1}\\
&+\left(\begin{array}{cc}
u_0 & u_1\mathrm{e}^{i\frac{2\pi}{3}\eta}\\
u_1\mathrm{e}^{-i\frac{2\pi}{3}\eta} & u_0\end{array}\right)\delta_{\bG,\bG'+(-1)^l\eta\left(\G_1+\G_2\right)},
\end{aligned}
\end{equation}
where $\bG,\bG'\in\{n_1\mathbf{G}_1+n_2\mathbf{G}_2\}$ define the lattices of mBZs, with $n_1$ and $n_2$ being integers, and a cutoff, $|\bG|$, $|\bG'|\leq6|\mathbf{G}_1|$, is applied. $\nu_\mathrm{F}$ is the Fermi velocity and $\hbar v_\mathrm{F} /(\sqrt{3}a)=2.37745$ eV. $\pmb{\sigma}^\eta = \left(\eta\sigma_x,\sigma_y\right)$, where $\sigma_x$ and $\sigma_y$ are Pauli matrices in sublattice space. The first part of $H^{s,\eta,\bk}_{\bG,\bG'}$ stands for intra-layer hopping of the top (1) layer and the bottom (2) layer, respectively, and the second part stands for inter-layer hopping  $u_0$ and $u_1$, which are actually the moir\'e potentials.

\begin{figure}[!ht]
\centering
\includegraphics[width=\linewidth]{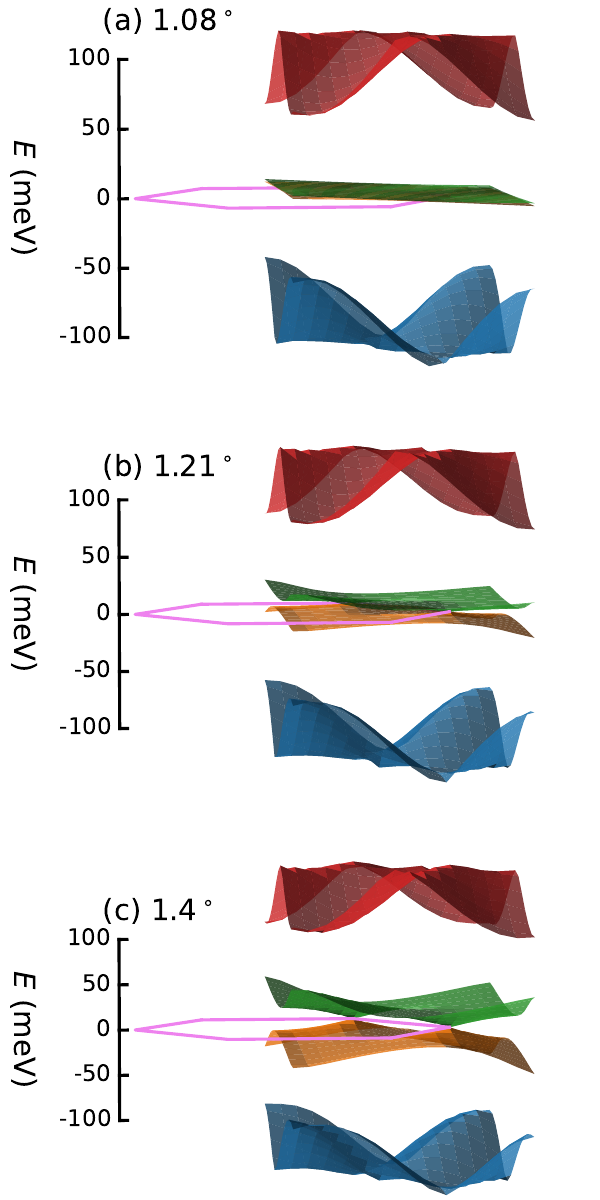}
\caption{\textbf{Four low-energy bands for $L=15$ without Coulomb interaction}. (a) 1.08$\degree$, (b) 1.21$\degree$, and (c) $1.4\degree$. Note that $u_0 = $ 60 meV, and the mBZ is denoted by the violet hexagon. The two low-energy bands are also shown in Fig.~\ref{fig:fig1} (a), (b) and (c) in the main text.}
\label{fig:fig6}
\end{figure}

The dispersion of $H_0$ for several twist angles are shown in Fig.~\ref{fig:fig6} with $u_0 = $ 60 meV, where enlarging of the bandwidth of two low-energy bands is seen with increasing $\Theta$ from MA, and the two low-energy bands are always isolated from the remote bands by $\sim$60 meV. For the cases considered in this study, the minimum gap between low-energy bands and remote bands is not less than 20 meV while the maximum Coulomb potential is less than 6 meV, as shown in Fig.~\ref{fig:V(0)}. Thus, the low-energy physics can be well captured by these two low-energy bands.

The single-gated Coulomb interaction after Fouriers transformation is
\begin{equation}
H_\mathrm{I}=\frac{1}{2\Omega}\sum_{\bq\in\mathrm{mBZ},\bG}V(\bq+\bG)\delta\rho_{\bq+\bG}\delta\rho_{-\bq-\bG},
\end{equation}
where
\begin{equation}
\begin{aligned}
&\delta\rho_{\bq+\bG}\\
&=\sum_{s,\eta,\bk,\bG',X'}
\left(d^\dagger_{s,\eta,\bk,\bG',X'}d_{s,\eta,\bk+\bq+\bG,\bG',X'}-\frac{\nu+4}{8}\delta_{\bq,0}\delta_{\bG,0}\right)\\
&=\sum_{s,\eta,\bk}
\left(\vd^\dagger_{s,\eta,\bk}\vd_{s,\eta,\bk+\bq+\bG}-\frac{\nu+4}{8}\delta_{\bq,0}\delta_{\bG,0}\right).
\end{aligned}
\end{equation}
Here $\delta\rho_{\bq+\bG}$ is the electron density operator with respect to $(\nu+4)/8$ filling, and $\nu$ is the filling parameter with $\nu = 0$ corresponding to the charge neutrality. $\Omega=N|\mathbf{a}_\mathrm{M1}||\mathbf{a}_\mathrm{M2}|\sqrt{3}/2$ is the total area of mori\'e superlattice in real space, while $\mathbf{a}_\mathrm{M1}$ and $\mathbf{a}_\mathrm{M2}$ are the lattice vectors of real-space unit cell of the superlattice with $|\mathbf{a}_{\mathrm{M1,M2}}|=\sqrt{3}a/[2\mathrm{sin}(\Theta/2)]$. For the notational simplicity, from here on $\bq+\bG$ is denoted as $\mQ$.
 
Denoting the eigenvalues and eigenvectors of the BM-model matrix $H^{s,\eta,\bk}$ in Eq.~\ref{eq:BM} as $\epsilon^{s,\eta}_{\bk,m}$ and $\left|u^{s,\eta}_{\bk,m}\right>$, $H^{s,\eta,\bk}\left|u^{s,\eta}_{\bk,m}\right>=\epsilon^{s,\eta}_{\bk,m}\left|u^{s,\eta}_{\bk,m}\right>$, or 
\begin{equation}
H^{s,\eta,\bk}U^{s,\eta,\bk}=U^{s,\eta,\bk}E^{s,\eta,\bk},
\end{equation}
with each column of $U^{s,\eta,\bk}$ consisting of $\left|u^{s,\eta}_{\bk,m}\right>$ and $E^{s,\eta,\bk}$ is a diagonal matrix consisting of $\epsilon^{s,\eta}_{\bk,m}$,
 then $H_0$ can be projected to the band basis as
\begin{equation}
\begin{aligned}
H_0=&\sum_{s,\eta,\bk}\vd_{s,\eta,\bk}^\dagger H^{s,\eta,\bk}\vd_{s,\eta,\bk}\\
=&\sum_{s,\eta,\bk}\vd_{s,\eta,\bk}^\dagger U^{s,\eta,\bk} {U^{s,\eta,\bk}}^\dagger H^{s,\eta,\bk}U^{s,\eta,\bk} {U^{s,\eta,\bk}}^\dagger\vd_{s,\eta,\bk}\\
=&\sum_{s,\eta,\bk}\vd_{s,\eta,\bk}^\dagger U^{s,\eta,\bk} E^{s,\eta,\bk} {U^{s,\eta,\bk}}^\dagger\vd_{s,\eta,\bk}.
\label{eq:proj_kinetics}
\end{aligned}
\end{equation}
Defining the projection
\begin{equation}
\vc_{s,\eta,\bk}^\dagger=\vd_{s,\eta,\bk}^\dagger U^{s,\eta,\bk},
\end{equation}
where $\vc_{s,\eta,\bk}$ is an operator vector with indices of $\bG$ and $X$, with the element, 
\begin{equation}
c^\dagger_{s,\eta,\bk,m}=\sum_{\bG,X}d^\dagger_{s,\eta,\bk,\bG,X}\left|u^{s,\eta}_{\bk,m}\right>_{\bG,X}, \end{equation}
then
\begin{equation}
\begin{aligned}
H_0=&\sum_{s,\eta,\bk}\vc_{s,\eta,\bk}^\dagger E^{s,\eta,\bk} \vc_{s,\eta,\bk}\\
=&\sum_{s,\eta,\bk,m}\epsilon^{s,\eta}_{\bk,m}c_{s,\eta,\bk,m}^\dagger c_{s,\eta,\bk,m}.
\label{eq:H0_methods}
\end{aligned}
\end{equation}
This is the projection from the plane-wave basis, $d^\dagger$, to the band basis, $c^\dagger$. To make the projection to two low-energy bands, $m$ relating to  these two bands are kept. 

Similarly for the interaction, 
\begin{equation}
\begin{aligned}
\vd^\dagger_{s,\eta,\bk}\vd_{s,\eta,\bk+\mQ}=&\vd^\dagger_{s,\eta,\bk}U^{s,\eta}_{\bk}{U^{s,\eta}_{\bk}}^\dagger U^{s,\eta}_{\bk+\mQ}{U^{s,\eta}_{\bk+\mQ}}^\dagger\vd_{s,\eta,\bk+\mQ}\\
=&\vc^\dagger_{s,\eta,\bk}{U^{s,\eta}_{\bk}}^\dagger U^{s,\eta}_{\bk+\mQ}\vc_{s,\eta,\bk+\mQ}\\
=&\sum_{m,n}c^\dagger_{s,\eta,\bk,m}\left({U^{s,\eta}_{\bk}}^\dagger U^{s,\eta}_{\bk+\mQ}\right)_{m,n}c_{s,\eta,\bk+\mQ,n}\\
=&\sum_{m,n}c^\dagger_{s,\eta,\bk,m}\lambda^{s,\eta}_{m,n}(\bk,\mQ)c_{s,\eta,\bk+\mQ,n},
\label{eq:proj_interaction}
\end{aligned}
\end{equation}
where the formfactor is defined as
\begin{equation}
\begin{aligned}
\lambda^{s,\eta}_{m,n}(\bk,\mQ)=&\left({U^{s,\eta}_\bk}^\dagger U^{s,\eta}_{\bk+\mQ}\right)_{m,n}\\
=&\langle u^{s,\eta}_{\bk,m}|u^{s,\eta}_{\bk+\mQ,n}\rangle.
\end{aligned}
\end{equation}
Note that we have made $U^{s,\eta}_{\bk} = U^{s,\eta,\bk}$ to make the expression compact. So,
\begin{equation}
\delta_\mQ=\sum_{s,\eta,\bk}\left(\sum_{m,n}\lambda^{s,\eta}_{m,n}(\bk,\mQ)c^\dagger_{s,\eta,\bk,m}c_{s,\eta,\bk+\mQ,n}-\frac{\nu+4}{8}\delta_{\bq,0}\delta_{\bG,0}\right)
\end{equation}
Moreover, since 
\begin{equation}
\begin{aligned}
\delta_{\bG,0}=&\sum_{m}\langle u^{s,\eta}_{\bk,m}|u^{s,\eta}_{\bk+\bG,m}\rangle\\
=&\sum_{m,n}\lambda^{s,\eta}_{m,n}(\bk,\bG)\delta_{m,n},
\end{aligned}
\end{equation}
and further,
\begin{equation}
\begin{aligned}
\delta_{\bq,0}\delta_{\bG,0}=\sum_{m,n}\lambda^{s,\eta}_{m,n}(\bk,\mQ)\delta_{\bq,0}\delta_{m,n},
\end{aligned}
\end{equation}
so that
\begin{equation}
\delta_\mQ=\sum_{s,\eta,\bk,m,n}\lambda^{s,\eta}_{m,n}(\bk,\mQ)\left(c^\dagger_{s,\eta,\bk,m}c_{s,\eta,\bk+\mQ,n}-\frac{\nu+4}{8}\delta_{\bq,0}\delta_{m,n}\right).
\end{equation}


Since the set of $\{|\mathcal{Q}|\neq 0\}$ is inverse symmetric, then
\begin{equation}
\begin{aligned}
H_\mathrm{I}&=\sum_{|\mathcal{Q}|\neq 0}\frac{1}{2\Omega}V(\mathcal{Q})\delta\rho_{\mathcal{Q}}\delta\rho_{-\mathcal{Q}}\\
&=\sum_{\bQ}\frac{1}{2\Omega}V(\bQ)(\delta\rho_{\bQ}\delta\rho_{-\bQ}+\delta\rho_{-\bQ}\delta\rho_{\bQ})\\
&=\sum_{\bQ}\frac{1}{4\Omega}V(\bQ)\left(\left(\delta\rho_{-\bQ}+\delta\rho_\bQ\right)^2-\left(\delta\rho_{-\bQ}-\delta\rho_\bQ\right)^2\right)\\
&=\sum_{\bQ}\frac{1}{4\Omega}V(\bQ)\left(A_\bQ^2-B_\bQ^2\right),
\end{aligned}
\label{eq:HI}
\end{equation}
where the set of $\{\bQ\}$ is any half of $\{|\mathcal{Q}|\neq0\}$. $H_0$ in Eq.~\eqref{eq:H0_methods} and $H_\mathrm{I}$ in Eq.~\eqref{eq:HI} are exactly the terms used in the Hamiltonian ~\eqref{eq:eq1} in the main text.

The KIVC order parameter is defined in the plane-wave basis as
\begin{equation}
\begin{aligned}
O_\mathrm{KIVC}(\bq)\equiv&\sum_{s,\bk}\vd^\dagger_{s,\bk+\bq}\eta_x\mu_y\sigma_x\vd_{s,\bk}\\
=&\sum_{s,\eta,\bk}\vd^\dagger_{s,\eta,\bk+\bq}\mu_y\sigma_x\vd_{s,-\eta,\bk},
\end{aligned}
\end{equation}
and likewise projected to the band basis as
\begin{equation}
\begin{aligned}
&O_\mathrm{KIVC}(\bq)\\
&=\sum_{s,\eta,\bk}\vd^\dagger_{s,\eta,\bk+\bq}U^{s,\eta}_{\bk+\bq}{U^{s,\eta}_{\bk+\bq}}^\dagger\mu_y\sigma_xU^{s,-\eta}_{\bk}{U^{s,-\eta}_{\bk}}^\dagger\vd_{s,-\eta,\bk}\\
&=\sum_{s,\eta,\bk}\vc_{s,\eta,\bk+\bq}^\dagger{U^{s,\eta}_{\bk+\bq}}^\dagger\mu_y\sigma_xU^{s,-\eta}_{\bk}\vc_{s,-\eta,\bk}\\
&=\sum_{s,\eta,\bk,m,n}c_{s,\eta,\bk+\bq,m}^\dagger\left({U^{s,\eta}_{\bk+\bq}}^\dagger\mu_y\sigma_xU^{s,-\eta}_{\bk}\right)_{m,n}c_{s,-\eta,\bk,n}\\
&=\sum_{s,\eta,\bk,m,n}c_{s,\eta,\bk+\bq,m}^\dagger\langle u^{s,\eta}_{\bk+\bq,m}|\mu_y\sigma_x|u^{s,-\eta}_{\bk,n}\rangle c_{s,-\eta,\bk,n}.
\end{aligned}
\end{equation}
Note that $\eta_x$ and $\mu_y$ are Pauli matrices for valleys and layers respectively. Similarly, the VP order parameter from plane-wave basis to band basis is
\begin{equation}
\begin{aligned}
O_\mathrm{VP}(\bq)\equiv&\sum_{s,\bk}\vd^\dagger_{s,\bk+\bq}\eta_z\mu_0\sigma_0\vd_{s,\bk}\\
=&-\sum_{s,\eta,\bk,m,n}\eta c_{s,\eta,\bk+\bq,m}^\dagger\langle u^{s,\eta}_{\bk+\bq,m}|\mu_0\sigma_0|u^{s,\eta}_{\bk,n}\rangle c_{s,\eta,\bk,n}\\
=&-\sum_{s,\eta,\bk,m,n}\eta c_{s,\eta,\bk+\bq,m}^\dagger\langle u^{s,\eta}_{\bk+\bq,m}|u^{s,\eta}_{\bk,n}\rangle c_{s,\eta,\bk,n}.
\end{aligned}
\end{equation}

\noindent{\textbf{Sec. II Gauge fixing}}

As can be seen from Eqs.~\eqref{eq:proj_kinetics} and Eq.~\eqref{eq:proj_interaction}, the projections from the plane-wave basis to the band basis are gauge-independent within each spin-valley flavor, so that it is unnecessary to fix gauges within each spin-valley flavor. For different valleys, we fix $\mathcal{PT}$, which is $\eta_x\mu_y\sigma_x$ in plane-wave basis or $\eta_yn_y$ in band basis with $n_y$ acting on bands. $\mathcal{PT}$ leads to complex conjugate between valleys.

In addition, to use the periodic condition, $c_{s,\eta,\bk+\bG,m}=c_{s,\eta,\bk,m}$, we impose
\begin{equation}
|u^{s,\eta}_{\bk+\bG,m}\rangle_{\bG',X}=|u^{s,\eta}_{\bk,m}\rangle_{\bG'-\bG,X}.
\end{equation}

\begin{figure}[!h]
\centering
\includegraphics[width=\linewidth]{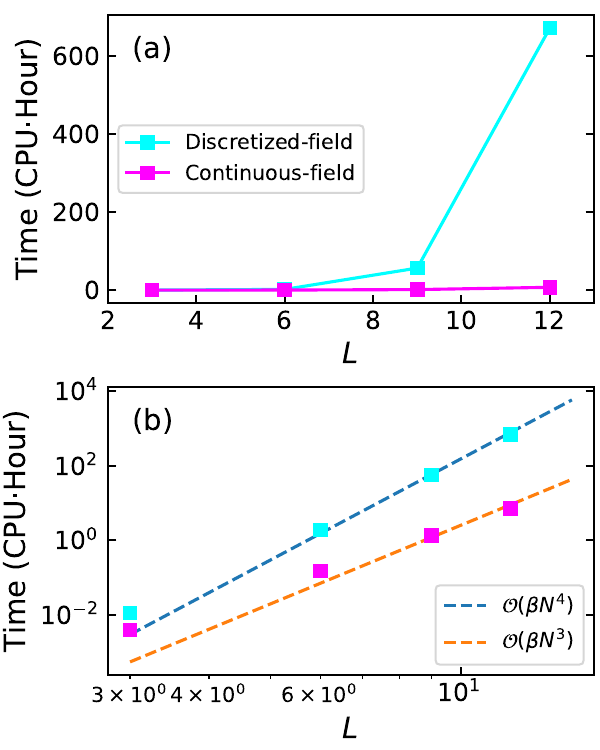}
\caption{\textbf{Computational complexities of DQMC and CFMC with the same physical settings}. We measure the CPU time consumed for 100 sweeps in the units of CPU core hours. Note that inverse temperature also sclaes with the systems size: $\beta\sim L$. The data can be very well described by the fits of $\beta N^4$ and $\beta N^3$ for DQMC and CFMC respectively, especially in the limit of large system sizes.}
\label{fig:fig7}
\end{figure}

\noindent{\textbf{Sec. III \ The CFMC algorithm}} 

The Hamiltonian in Eq.~\eqref{eq:eq1} has been solved with momentum-space QMC method with discretized auxiliary fields~\cite{zhangMomentum2021,hofmannFermionic2022}. Similar to other determinantal QMC methods, it applies the Hubbard-Stratonovich transformation to decouple the Coulomb interaction, following the Trotter decomposition~\cite{BlankenbeclerMonte1981,WhiteNumerical1989,AssaadWorld-line2008,bercx2017alf,HirschDiscrete1983}.
Due to $\mathcal{PT}$ symmetry the system at charge neutrality is free of sign problem~\cite{zhangMomentum2021,hofmannFermionic2022}. The simulation temperature is set to be $1/T\sim L$ with $T =$ 0.5 meV for $L =$ 6.

Original version of the momentum-space QMC method~\cite{zhangMomentum2021,panDynamical2022} uses the discrete version of the Hubbard-Stratonovich transformation following a Trotter decomposition:
\begin{equation}
\begin{aligned}
\exp\left(-\alpha_1(\bQ) A_\bQ^2\right)=&\frac{1}{4}\sum_{l_{\bQ,1}}\gamma\left(l_{\bQ,1}\right)\exp\left(\i\xi\left(l_{\bQ,1}\right)\sqrt{\alpha_1(\bQ)}A_\bQ\right)\\
&+\mathcal{O}\left({\alpha_1(\bQ)}^4\right),
\label{eq:eq3}
\end{aligned}
\end{equation}
and
\begin{equation}
\begin{aligned}
\exp\left(\alpha_1(\bQ) B_\bQ^2\right)=&\frac{1}{4}\sum_{l_{\bQ,2}}\gamma\left(l_{\bQ,2}\right)\exp\left(\xi\left(l_{\bQ,2}\right)\sqrt{\alpha_1(\bQ)}B_\bQ\right)\\
&+\mathcal{O}\left({\alpha_1(\bQ)}^4\right).
\end{aligned}
\label{eq:eq4}
\end{equation}
Here $l_{\bQ,1(2)}=\pm1,\pm2$ is a discrete auxiliary field, and the coeficients, $\gamma (\pm 1)=1+\sqrt{6}/3$, $\gamma (\pm 2)=1-\sqrt{6}/3$, $\xi (\pm 1)=\pm\sqrt{6-2\sqrt{6}}$, and $\xi (\pm 2)=\pm\sqrt{6+2\sqrt{6}}$ are taken from the eighth-order approximation as in ~\cite{AssaadWorld-line2008}. $\alpha_1(\bQ)=\triangle\tau V(\bQ)/4\Omega<8\times10^{-3}$ with the imaginary time $\beta=1/(k_\mathrm{B}T)$ evenly sliced into $N_\tau$ pieces, as $\triangle\tau=\beta/N_\tau$. The partition function, $Z=\Tr\{\exp\left(-\beta H\right)\}$, yields
\begin{equation}
Z=\sum_{\{l_{\tau,\bQ,1},l_{\tau,\bQ,2}\}}\Tr \ \{U_C\}
\label{eq:eq5}
\end{equation}
with
\begin{equation}
\begin{aligned}
U_C=&\prod_{\tau=\triangle\tau}^\beta\exp\left(-\triangle\tau H_0\right)\prod_\bQ\frac{1}{16}\gamma\left(l_{\tau,\bQ,1}\right)\gamma\left(l_{\tau,Q,2}\right)\\
&\times\exp\left(\i\xi\left(l_{\tau,\bQ,1}\right)\sqrt{\alpha_1(\bQ)}A_\bQ\right)\\
&\times\exp\left(\xi\left(l_{\tau,\bQ,2}\right)\sqrt{\alpha_1(\bQ)}B_\bQ\right)\\
&+\mathcal{O}(8\times10^{-3}),
\end{aligned}
\end{equation}
where $C$ stands for a configuration of the auxiliary fields, and $\mathcal{O}(8\times10^{-3})$ stands for an error less than $8\times10^{-3}$, usually $5\times10^{-3}$, of the exact value.

With the partition function at hand, measurements can be done by sampling the discrete auxiliary fields with the Metropolis-Hastings scheme. Local update of one auxiliary field $l_{\bQ,1(2)}$ yields an acceptance rate around 0.4 and cluster update of multiple fields yields a vanishing accepting rate. Each local update of a field $l_{\tau,\bQ,1(2)}$ leads to the computation of matrix determinant with the complexity $\mathcal{O}(N^3)$, due to that $A_\bQ$ or $B_\bQ$ is non-diagonal. Combining with the absence of the efficient cluster update scheme for discrete fields, it leads to an overall complexity of at least $\mathcal{O}(\beta N^4)$ in the original momentum-space QMC  for a sweep to update $2N_\tau N_\bQ$ fields. Note that the number of the combinations $(\bQ,1(2))$ is $2N_\bQ\approx3N$ (the cutoff at $|\mathbf{G}_1|$ makes the number of momentum exchanges roughly 3 times of the number of $\bk$ points in mBZ).
Therefore, the previous studies on TBG by QMC were heavily constrained by the accessible system sizes with the largest one being  $L=9$~\cite{hofmannFermionic2022,panDynamical2022}. 
 
To reduce the computational complexity, we develop a continuous-field Monte Carlo (CFMC) scheme to carry out cluster updates of  multiple auxiliary fields at once. First of all, we switch to continuous auxiliary fields using the Gaussian integrals
\begin{equation}
\e^{\frac{1}{2}\alpha_2A^2}=\frac{1}{\sqrt{2\pi}}\int^\infty_{-\infty}d\phi\e^{-\frac{1}{2}\phi^2}\e^{-\phi\sqrt{\alpha_2}A},
\end{equation}
to rewrite the partition function of  Eq.~\eqref{eq:eq5} in the form
\begin{equation}
\begin{aligned}
Z=&\int\left(\prod_{\tau,\bQ}d\phi_{\tau,\bQ,1}d\phi_{\tau,\bQ,2}\right)\e^{-\frac{1}{2}\sum_{\tau,Q}\left(\phi^2_{\tau,\bQ,1}+\phi^2_{\tau,\bQ,2}\right)}\\
&\times\Tr\left(\prod_\tau\e^{\i\sum_\bQ\left(-\phi_{\tau,\bQ,1}\sqrt{\alpha_2(\bQ)}A_\bQ+\i\phi_{\tau,\bQ,2}\sqrt{\alpha_2(\bQ)}B_\bQ\right)}\e^{-\triangle\tau H_0}\right)\\
&+\mathcal{O}(8\times10^{-3}),
\end{aligned}
\label{eq:zcfmc}
\end{equation}
where $\alpha_2(\bQ)=\triangle\tau V(\bQ)/2\Omega$. Note that the Gaussian decoupling is exact, so that the error of $Z$ here comes from the Trotter decomposition, same as in Eq.~\ref{eq:eq5}.

We employ the same Metropolis-Hastings scheme for the updates of the continuous fields $\phi_{\tau,\bQ,1(2)}$. Small updates of $\phi_{\tau,\bQ,1(2)}$ would make an acceptance rate around one while large updates would lead to a vanishing acceptance rate. The changes of $\phi_{\tau,\bQ,1(2)}$, $\triangle\phi_{\tau,\bQ,1(2)}$, can be proposed from a Gaussian distribution or from the Hamiltonian's dynamics. 

\vspace{0.4cm}
\noindent{\textbf{Sec. III A \ The CFMC algorithm: Gaussian Update Scheme}}
\vspace{0.4cm}

For the Gaussian proposal, we tune the magnitudes of $\triangle\phi_{\tau,\bQ,1(2)}$ to make the acceptance rate around 0.4. The difference between the new and the old value of the auxiliary field is written as  $\triangle\phi_{\tau,\bQ,1(2)}=rx$ with $r$ being the control parameter to tune the magnitudes of $\triangle\phi_{\tau,\bQ,1(2)}$ and $x$ generated from a standard normal distribution, $\mathcal{N}(0,1)$. The $r$ is updated by  $r\rightarrow r+(R_\mathrm{a}-0.4)\triangle r$ where $\triangle r$ is a positive constant and $R_\mathrm{a}$ is the measured acceptance rate. $R_\mathrm{a}$ smaller than 0.4 means $\triangle\phi_{\tau,\bQ,1(2)}$ are larger than the desirable ones, and thus $r$ is decreased, and vise versa. $R_\mathrm{a}\sim0.4$ should be achieved before the Markov process reaches equilibrium distribution. 

All $\phi_{\tau,\bQ,1(2)}$ of a specific imaginary time $\tau$, with the number of $\approx3N$ in total, are updated each time. Thus, only one calculation of matrix determinant with complexity $\mathcal{O}(N^3)$ is needed for each imaginary time $\tau$.
 
 The full protocol of CFMC with Gaussian proposals can be described as follows 
\begin{enumerate}
\item Generating all $\phi_{\tau,\bQ,1(2)}$ fields using normal distribution $\mathcal{N}(0,1)$ and the current value of $r$ parameter (initially we set $r$ and $\Delta r$ to 0.1).
\item Updating all $\phi_{\tau,\bQ,1(2)}$ of $\tau=\beta$ by $\triangle\phi_{\tau,\bQ,1(2)}=rx$ and accept-rejecting the update according to Metropolis-Hastings scheme. 
\item Repeating the updates for all time slices from $\beta$ to $\triangle\tau$ then reversely to complete a full sweep, in the process of which $R_\mathrm{a}$ is calculated. After each sweep altering $r$ using the rule $r\rightarrow r+(R_\mathrm{a}-0.4)\triangle r$.
\item Repeating the sweep until sufficient statistics is gained for all observables. 
\end{enumerate}

Since the auxiliary fields are continuous, it is easy to adopt more sophisticated updating schemes like Hamiltonian dynamics to improve the proposed $\triangle\phi_{\tau,\bQ,1(2)}$.

\vspace{0.4cm}
\noindent{\textbf{Sec. III B The CFMC algorithm: Hamiltonian Update Scheme}}
\vspace{0.4cm}

One way to enhance the sampling of continuous fields is by using Hamiltonian dynamics to update the fields, rather than relying on simple Gaussian proposals. This approach can significantly boost the acceptance rate, even for large global updates — a key feature of the so-called Hybrid Monte Carlo (HMC) algorithm. This method have been already applied in condensed matter systems, especially those with long range Coulomb interactions  ~\cite{buividovichNumerical2019}.

The partition function in Eq.~\ref{eq:zcfmc} can be further expressed as 
\begin{equation}
\begin{aligned}
Z=&\int\left(\prod_{i}d\phi_{i}d p_{i}\right)\e^{-\frac{1}{2}\sum_{i}\left(\phi^2_{i}+p^2_{i}\right)}\det\left(\I+B_C(\beta,0)\right)\\
&+\mathcal{O}(8\times10^{-3}),
\end{aligned}
\end{equation}
where and the index $i\equiv({\tau,\bQ,m})$ with $m=1,2$ is employed for compactness and an artificial momenta $p_{i}$ is introduced for $\phi_{i}$. Moreover,
\begin{equation}
\begin{aligned}
&\det\left(\I+B_C(\beta,0)\right)=\\
&\Tr\left(\prod_\tau\e^{\i\sum_\bQ\left(-\phi_{\tau,\bQ,1}\sqrt{\alpha_2(\bQ)}A_\bQ+\i\phi_{\tau,\bQ,2}\sqrt{\alpha_2(\bQ)}B_\bQ\right)}\e^{-\triangle\tau H_0}\right),
\end{aligned}
\end{equation}
which is nonnegative for charge-neutral TBG as mentioned before. Furthermore, the partition function can be expressed as
\begin{equation}
\begin{aligned}
Z=&\int\left(\prod_{i}d\phi_{i}d p_{i}\right)\e^{-\mathcal{ H}}+\mathcal{O}(8\times10^{-3}),
\end{aligned}
\end{equation}
with the effective Hamiltonian 
\begin{equation}
\mathcal{H}\equiv\frac{1}{2}\sum_{i}\left(\phi^2_{i}+p^2_{i}\right)-\ln \left(\det\left(\I+B_C(\beta,0)\right)\right).
\end{equation}

The Hamiltonian dynamics is applied as  
\begin{equation}
\begin{aligned}
\frac{dp_{i}}{dt}=&-\frac{\partial\mathcal{H}}{\partial\phi_{i}}\\
= &-\phi_{i}+4\Tr\left(J_\bQ(\I-G(\tau))\right) +\mathcal{O}(10^{-2}) \\
\frac{d\phi_{i}}{dt}=&\frac{\partial\mathcal{H}}{\partial p_{i}}\\
=&p_{i},
\label{eq:ham_dyn}
\end{aligned}
\end{equation}
where $G(\tau)$ is the equal-time Green's function for $s=\uparrow$ and $\eta=+$, and $J_\bQ$ is the matrix form of $A_\bQ$ or $B_\bQ$  for $m = $ 1 or 2. The Hamiltonian dynamics facilitates the increased acceptance rates through the conservation of the Hamiltonian along the equal-energy trajectory. 


\vspace{0.4cm}
\noindent{\textbf{Sec. III C The outperformance of the CFMC algorithm}}
\vspace{0.4cm}

Comparison of performances of CFMC and discretized-field QMC (DQMC) algorithms is shown in Fig.~\ref{fig:fig7} using the same physical settings. We can clearly see that CFMC indeed reduces the computational complexity from $O(\beta N^4)$ to $O(\beta N^3)$, since the CPU time consumed by DQMC grows much faster than the one by CFMC, as system size increases. 

The CPU time per update is not the only metric which defines the performance of the Monte Carlo scheme. It can be that different updates lead to substantially different distributions of the observables (despite the average values being fixed) \cite{PhysRevE.106.025318}. Thus, the faster algorithm can potentially suffer from larger statistical fluctuations, which can even cancel the gain in the speed of update. In order to check this, we plot the Monte Carlo time series of an IVC structure factor from DQMC and CFMC simlations. Results for  $\Theta=1.08\degree$, $1.2\degree$, and $1.3\degree$ are shown in Fig.~\ref{fig:fig12}. The histograms for the time series are drawn on top of them using the red and blue lines, respectively. The overlap of histograms indicates that fluctuations of DQMC and CFMC are very similar, thus the overall increase in performance from $O(\beta N^4)$ scaling to $O(\beta N^3)$ sclaing still holds.

Finally, we also compare autocorrelation for the CFMC and DQMC methods (see Fig.~\ref{fig:fig13}). Since the measurements are done after the updates of the fields in each time slice, we compare autocorrelation for measurements separated by a Euclidean time interval $\Delta t$. The autocorrelation time is indeed larger for CFMC, which offsets the factor of $N$ in speedup in simulations at MA and at the smallest lattice size. However, since the autocorrelation time does not depend on the lattice size, moving away from $L=6$ and $\Theta = 1.08^\degree$ reveals a greater advantage: the difference in scaling ($O(\beta N^4)$ for DQMC versus $O(\beta N^3)$ for CFMC) quickly outweighs the impact of the increased autocorrelation time (see the concrete examples in the caption for the Fig.~\ref{fig:fig13}). We therefore conclude that CFMC significantly enhances the performance of the Monte Carlo simulations.

In addition, we note that our CFMC algorithm can be applied not only to TBG but also to other systems with long-range interactions where the local update of discrete auxiliary field is at least $O(N)$ times more expensive due to the coupling of auxiliary filed to fermionic bilinear with dense matrix in momentum or configuration space. One example of another system is the projected half-filled Landau level which maps to the (2+1)D SO(5) nonlinear sigma model with Wess-Zumino-Witten term~\cite{chenPhases2024,chen2024emergent,ippolitiHalf2018,wangPhase2021,leeWess2015}.

\vspace{0.5cm}
\noindent{\textbf{Sec. IV \ Stochastic analytic continuation}} 

The stochastic analytic continuation (SAC)~\cite{Sandvik1998Stochastic,beachIdentifying2004,panDynamical2022,zhangSuperconductivity2021,panThermodynamic2023,huangEvolution2024,shao2023progress} method is also employed to extract the real frequency spectral functions. The Green's function relates to the real frequency spectral function as ~\cite{panDynamical2022}
\begin{equation}
    G(\bk,\tau)=\int^\infty_{-\infty}d\omega\left(\frac{\mathrm{e}^{-\omega\tau}}{1+\mathrm{e}^{-\beta\omega}}\right)A(\bk,\omega).
\label{eq:GA}
\end{equation}
A variational ansatz of $A(\bk,\omega)$ simulating an annealing process is applied.  For each momentum $\bk$, $A(\bk,\omega)=\sum_{i=1}^{N_\omega}A_i\delta(\omega-\omega_i)$ where frequencies $\omega_i$ initially form regular grid with length of $0.5k_\mathrm{B}T$ of $N_\omega=4000$ points symmetrically distributed around 0. Then $A(\bk,\omega)$ is estimated by sampling over the grid of $\omega_i$,  using Eq. ~\ref{eq:GA} and the effective action
\begin{equation}
\begin{aligned}
    \chi^2=&\sum_{\tau_1,\tau_2}\left(\overline{G}(\tau_1)-\int_{-\infty}^\infty d\omega\left(\frac{\mathrm{e}^{-\omega\tau}}{1+\mathrm{e}^{-\beta\omega}}\right)A(\omega)\right)(C^{-1})_{\tau_1,\tau_2}\\
    &\times\left(\overline{G}(\tau_2)-\int_{-\infty}^\infty d\omega\left(\frac{\mathrm{e}^{-\omega\tau}}{1+\mathrm{e}^{-\beta\omega}}\right)A(\omega)\right),
\end{aligned}
\end{equation}
where
\begin{equation}
\begin{aligned}
    &C_{\tau_1,\tau_2}\\
    &=\frac{1}{N_b(N_b-1)}\sum_{b=1}^{N_b}\left(G^b(\tau_1)-\overline{G}(\tau_1)\right)\left(G^b(\tau_2)-\overline{G}(\tau_2)\right),
\end{aligned}
\end{equation}
with $\tau_1$ and $\tau_2$ referring to selected imaginary times where the errors of Green's functions are less than 0.1. $G^b(\tau_{1,2})$ is the Green's function of $b$-th bin at $\tau_{1,2}$, and $\overline{G}(\tau_{1,2})$ is the average of $G^b(\tau_{1,2})$ over all bins. Importance sampling of Metropolis-Hastings type is employed and the sampling weight is $W=\exp\left(-\chi^2/(2T_1)\right)$ where $T_1$ is the artificial temperature. According to the general scheme of the stochastic analytical continuation, it decreases from 5 by a factor of 0.6 in each step. The final value of $T_1$ is defined by the average effective action satisfying the relation$\langle\chi^2\rangle=\chi_\mathrm{min}^2+2\sqrt{\chi_\mathrm{min}^2}$~\cite{panDynamical2022}, and this final value of $T_1$ is used to obtain the spectra.

\vspace{0.5cm}
\noindent{\textbf{Sec. V \ Computation of the free energy}} 

The free energy is difficult to obtain numerically, since exponentially large or small numbers are usually involved. Let us denote the partition function as $Z=\sum_CW_CP_C$, where $W_C$ corresponds to the bosonic part of the action and $P_C$ corresponds to the fermionic determinant. Since the kinetic energy is included as $\e^{-\triangle\tau H_0}$ in Eq.~\ref{eq:zcfmc}, $P_C$ becomes exponentially large as the kinetic energy grows with twist angle (see Fig.~\ref{fig:fig6}). One might think that positive and negative terms in the matrix $\e^{-\triangle\tau H_0}$ form pairs, so that they cancel each other and the magnitude of $P_C$ does not change a lot with the twist angle. However, the identity matrix in the relation, $\mathrm{Tr}\left(\e^{\sum_{i,j}c^\dagger_iA_{i,j}c_j}\e^{\sum_{i,j}c^\dagger_iB_{i,j}c_j}\right)=\det\left(\mathrm{I}+\e^A\e^B\right)$, forbids this cancellation. 

The free energy $F$ can be theoretically measured by 
\begin{equation}
    \begin{aligned}
        F&\equiv-\frac{1}{\beta}\ln(Z)\\
        &=-\frac{1}{\beta}\ln\left(\sum_CW_CP_C\right)\\
        &=-\frac{1}{\beta}\ln\left(\left(\sum_CW_C\right)\frac{\sum_CW_CP_C}{\sum_CW_C}\right)\\
        &=-\frac{1}{\beta}\ln\left(\sum_CW_C\right)-\frac{1}{\beta}\ln\left(\frac{\sum_CW_CP_C}{\sum_CW_C}\right),
    \end{aligned}
\end{equation}
which simply means the importance sampling of $P_C$ over the weight defined by $W_C$. However, $P_C$ is exponentially large in our case, so that direct sampling of a moderate-size system is inaccessible. 

Recently, an integral algorithm to efficiently sample the exponential observable such as the free energy and entanglement entropy was developed ~\cite{zhangIntegral2024}. Dumping the constant number $-\ln\left(\sum_CW_C\right)/\beta$ for a specific $T$, we can rewrite $F$ as \begin{equation}
    F=-\frac{1}{\beta}\int_0^1dt\frac{\partial\ln(f(t))}{\partial t},
    \label{eq_F}
\end{equation}
if $f(1)=\left(\sum_CW_CP_C\right)/\left(\sum_CW_C\right)$ and $f(0)=1$. Subsequently, we define $f(t)$ as $f(t)=\langle\e^{tX}\rangle$, so that $\langle\e^{X}\rangle=\left(\sum_CW_CP_C\right)/\left(\sum_CW_C\right)$ and $X=\ln(P)$. Substituting it in \eqref{eq_F}, we get 
\begin{equation}
\begin{aligned}
    F=&-\frac{1}{\beta}\int_0^1dt\frac{\langle X\e^{tX}\rangle}{\langle\e^{tX}\rangle}\\
    =&-\frac{1}{\beta}\int_0^1dt\frac{\sum_CW_CP_C^t\ln(P_C)}{\sum_CW_CP_C^t}.
\end{aligned}
\end{equation}
Thus, the exponentially large values of observable are avoided and we need only the importance samplings of $\ln(P_C)$ over the weight of $W_CP_C^t$.\\

\noindent{\textbf{Sec. VI \ Pad\'e approximants and scaling exponents.}}

The critical exponents characterize a particular universality class and depend on the dimensionality, symmetry and the relevant degrees of freedom. The transition to the KIVC order in TBG falls in the universality classe of the chiral XY GN model. This was considered previously via perturbative renormalisation up to four-loop order in an $\epsilon$ expansion 
\cite{PhysRevD.96.096010}, where $\epsilon$ is the deviation from the the upper critical spacetime dimension $D=4-\epsilon$ of the model In Ref. 
\cite{PhysRevD.96.096010}, a scale dependence is introduced in the fields $\phi$ and $\psi$, as well as for the mass term $m$, quartic coupling $\lambda$, and Yukawa coupling $g$ from Eq.~(\ref{eqn:GNYLag}), and their renormalization is calculated. Based on this, the inverse correlation-length exponent $\nu^{-1}$, the correction-to-scaling exponent $\omega$, the fermion and the boson anomlous dimension $\eta_\psi,\eta_{\phi}$ are determined and their expression for general fermion flavour number $N_f$ is provided up to fourth order in $\epsilon$. We use these results and apply them to our case of interest, the KIVC-DSM transition in TBG. To obtain an estimate for the critical exponents for the case of (2+1) dimensions, we calculate the different Pad\'e approximants 
\begin{equation}
    P_{[m/n]}(\epsilon) = \frac{a_0 + a_1\epsilon + \dots + a_m\epsilon^m}{1 + b_1\epsilon + \dots + b_n\epsilon^n}
\end{equation}
for a general fermion flavor number $N_f$ for the  exponents $\nu^{-1},\eta_{\phi}$. Selecting only those that do not exhibit any singularities or poles, which in our case are $P_{[2/2]}(\epsilon)$ and $P_{[3/1]}(\epsilon)$, we obtain the numerical values for $N_f=16$ and $\epsilon=1$ and use the hyperscaling relation to determine the order parameter exponent $\beta$, $\beta=\nu{(D-2+\eta_{\phi})}/{2}$. We subsequently average over these values and based on those provide the numerical uncertainties, and follow the same procedure to determine the correction-to-scaling exponents $\omega$.\\



\bibliography{ref}
\vspace{1cm}

\noindent{\textbf{\large Acknowledgments}}\\
We thank Michael Scherer for valuable input on the critical exponents from $4-\epsilon$ expansion and acknowledge discussions with Lukas Janssen on similar topics. C.H. and Z.Y.M. acknowledge the support from the
Research Grants Council (RGC) of Hong Kong Special Administrative Region (SAR) of China (Project Nos. 17301721, AoE/P701/20, 17309822, C7037-22GF, 17302223, 17301924), the ANR/RGC Joint Research Scheme sponsored by RGC of Hong Kong and French National Research Agency (Project No. A HKU703/22) and the HKU Seed Funding for Strategic Interdisciplinary Research. L.C. was funded by the European Union (ERC-2023-STG, Project 101115758 - QuantEmerge). Views and opinions expressed are, however, those of the authors only and do not necessarily reflect those of the European Union or the European Research Council Executive Agency. Neither the European Union nor the granting authority can be held responsible for them. M.U. and F.F.A. thank  the  DFG   for financial support  under the projects AS120/19-1, Project number 530989922. 
F.F.A acknowledges financial support from the DFG through the W\"urzburg-Dresden Cluster of Excellence on Complexity and Topology in Quantum Matter - \textit{ct.qmat} (EXC 2147, Project No.\ 390858490)   as  well as  the SFB 1170 on Topological and Correlated Electronics at Surfaces and Interfaces (Project No.\  258499086). We thank HPC2021 system under the Information Technology Services and the Blackbody HPC system at the Department of Physics, University of Hong Kong, as well as the Beijng PARATERA Tech CO., Ltd. (URL: https://cloud.paratera.com) for providing computing resources that have contributed to the research results reported within this paper.
\\




\clearpage
\onecolumngrid

\begin{center}
	\textbf{\large Supplemental Material for \\"Angle-Tuned Gross-Neveu Quantum Criticality in Twisted Bilayer Graphene: a Quantum Monte Carlo study"}
\end{center}
\section{Maximum of the single-gated screened Coulomb interactions}
\begin{figure}[!h]
\centering
\includegraphics[width=0.8\linewidth]{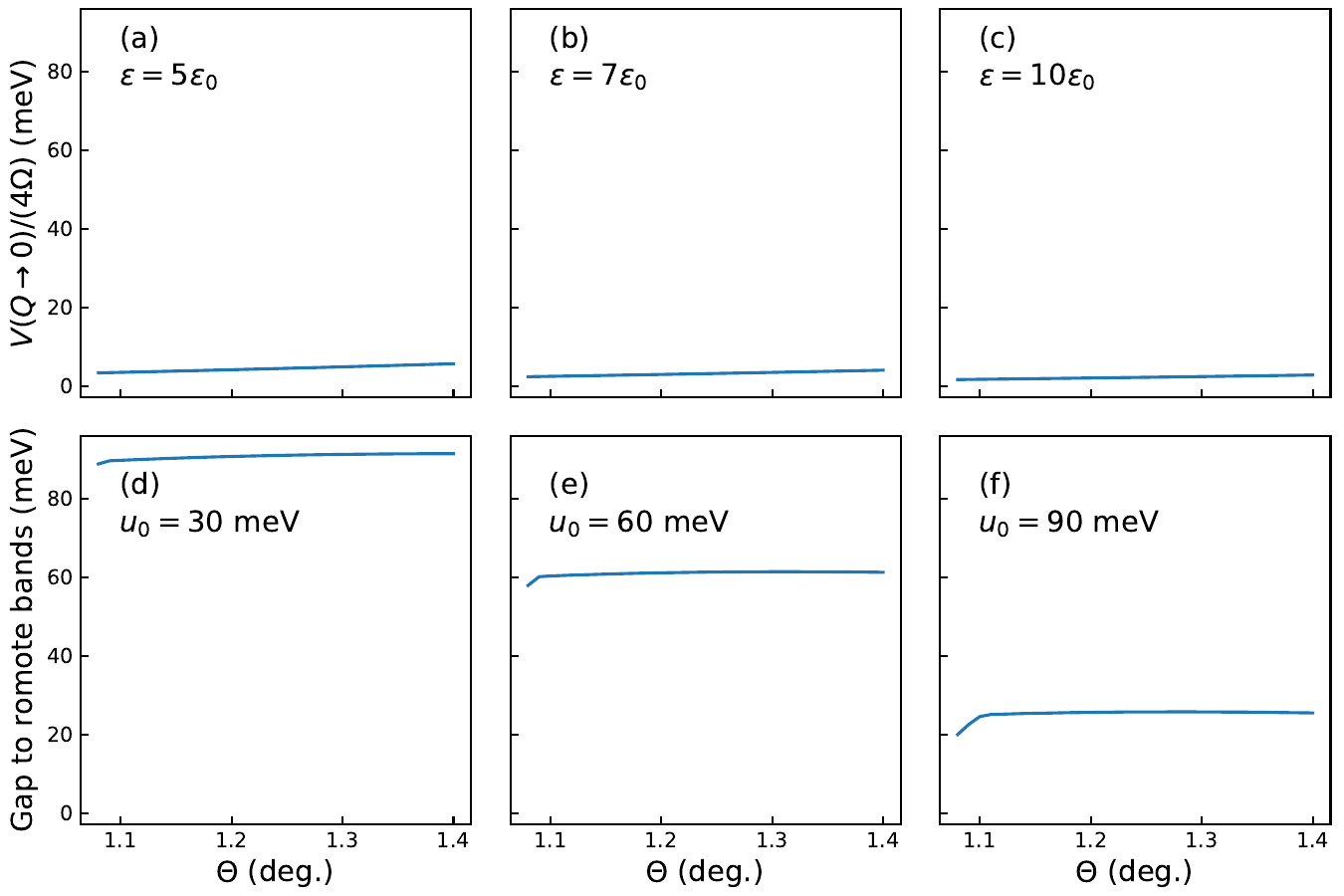}
\caption{(a) - (c) The maximum strengths of the screened Coulomb potentials as functions of the twist angle with different permittivities. (d)-(f) The gaps from two low-energy bands to remote bands of BM model with various $u_0$. Note that from their expressions, Coulomb potentials are irrelevant of $u_0$, and the gaps are independent of $\varepsilon$.}
\label{fig:V(0)}
\end{figure}

From its expression,
\begin{equation}
\frac{V(\bQ)}{(4\Omega)}=\frac{e^2}{8\Omega \varepsilon|\bQ|}\left(1-\mathrm{e}^{-|\bQ| d}\right),
\end{equation}
the maximum value of $V(\bQ)/(4\Omega)$ is at $\bQ\rightarrow0$. In Fig.~\ref{fig:V(0)} (a) - (c), the maxima of $V(\bQ)/(4\Omega)$ for the three permittivities are shown as functions of twist angle. Smaller $\varepsilon$ leads to stronger Coulomb interactions with the maximum at 1.4$\degree$ for $\varepsilon = 5\varepsilon_0$ among the cases considered in this study. This maximum is less than 6 meV. The gaps from two low-energy bands to remote bands of BM model are shown in Fig.~\ref{fig:V(0)} (d)-(f) respectively for $u_0 = $ 30, 60, and 90 meV, which indicate the gap for a specific $u_0$ is quite stable versus twist angle. It is clearly that the maximun of the Coulomb interactions are always much less than the minimum of the gaps, which is not less than 20 meV, so the projections to two low-energy bands are valid for the cases considered in this study.

\section{Altering $u_0$}
\begin{figure*}[!ht]
\centering
\includegraphics[width=\linewidth]{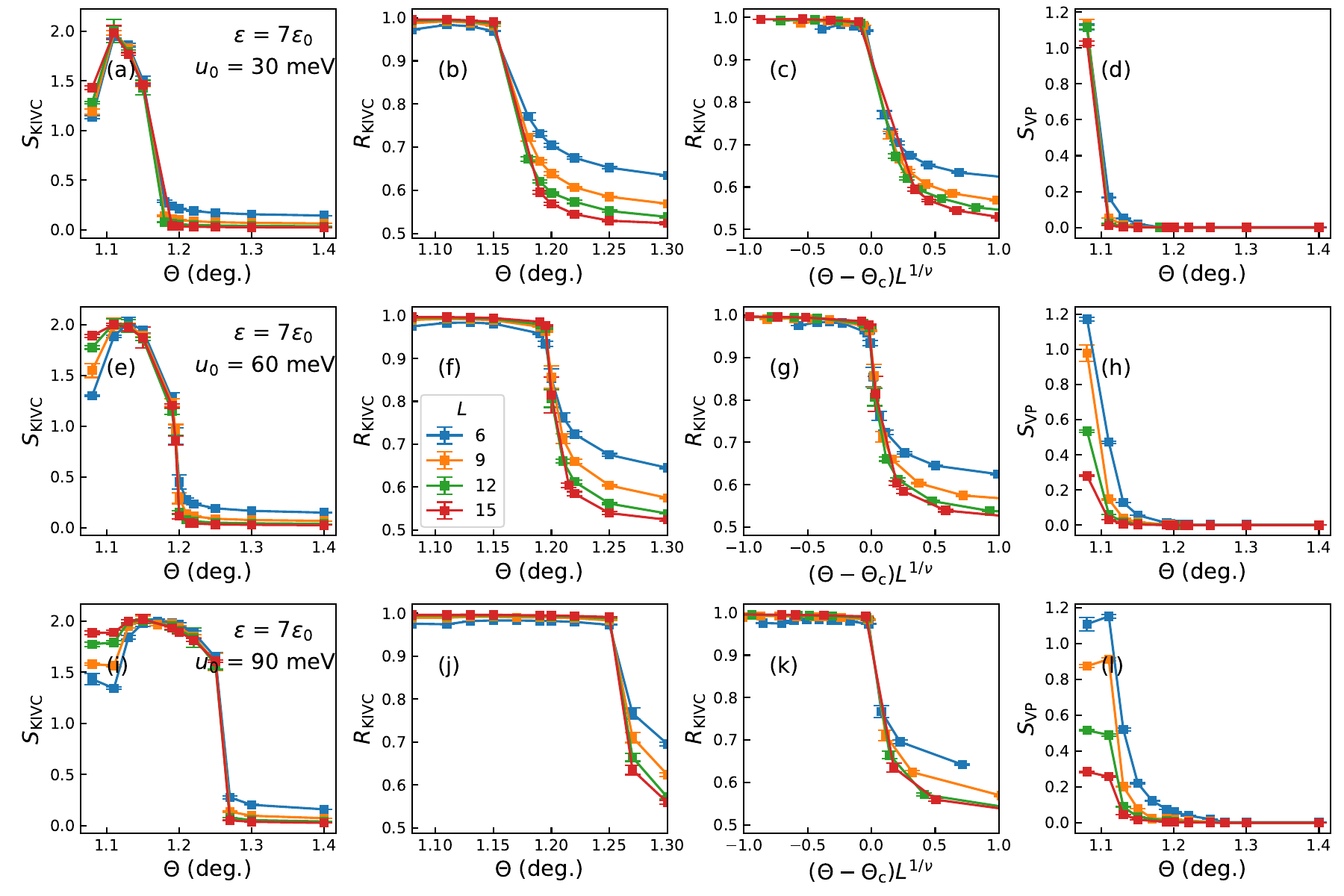}
\caption{ \textbf{The GN-QCP analysis from the KIVC order with different $u_0$}. Here $\varepsilon=7\varepsilon_0$. For $u_0 = $ 30, 60, and 90 meV, respectively, each row contains the structure factor $S$ of the KIVC order, its correlation ratio, finite-size collapse, and $S$ of the VP order, on the lattices with $L=6, 9, 12$, and 15 as functions of $\Theta$. We set $\beta \propto L$ in these CFMC simulations. From $1.08\degree$ to $1.4\degree$, the KIVC order continuously vanishes (although there are strong competition from VP when $\Theta<1.15\degree$) as the systems evolve from KIVC insulator to DSM. Correlation ratio gives the critical angle, $\Theta_{\c}=1.159\degree, 1.197\degree, \mathrm{and} \ 1.254\degree$, respectively for $u_0 = $ 30, 60, and 90 meV. The finite-size data collapses are consistent to the critical exponent, $\nu=1.13$ of the $N_f=16$ GN-QCP.}
\label{fig:R1}
\end{figure*}

Here we provide the outcomes on the correlations of KIVC and VP from another two cases of $u_0 = $ 30 and 90 meV when fixing $\varepsilon=7\varepsilon_0$, as shown in Fig.~\ref{fig:R1}. Both cases show similar GN-QCPs, judging from the correlations and data collapses, as also discussed in the main text. Note that a larger $u_0$ leads to a larger critical angle, with the $\Theta_{\c} = $1.159$\degree$, 1.197$\degree$, and 1.254$\degree$, respectively for $u_0 = $30, 60, and 90 meV.

\section{Comparison of the critical exponents}

\begin{figure}[!ht]
\centering
\includegraphics[width=0.8\linewidth]{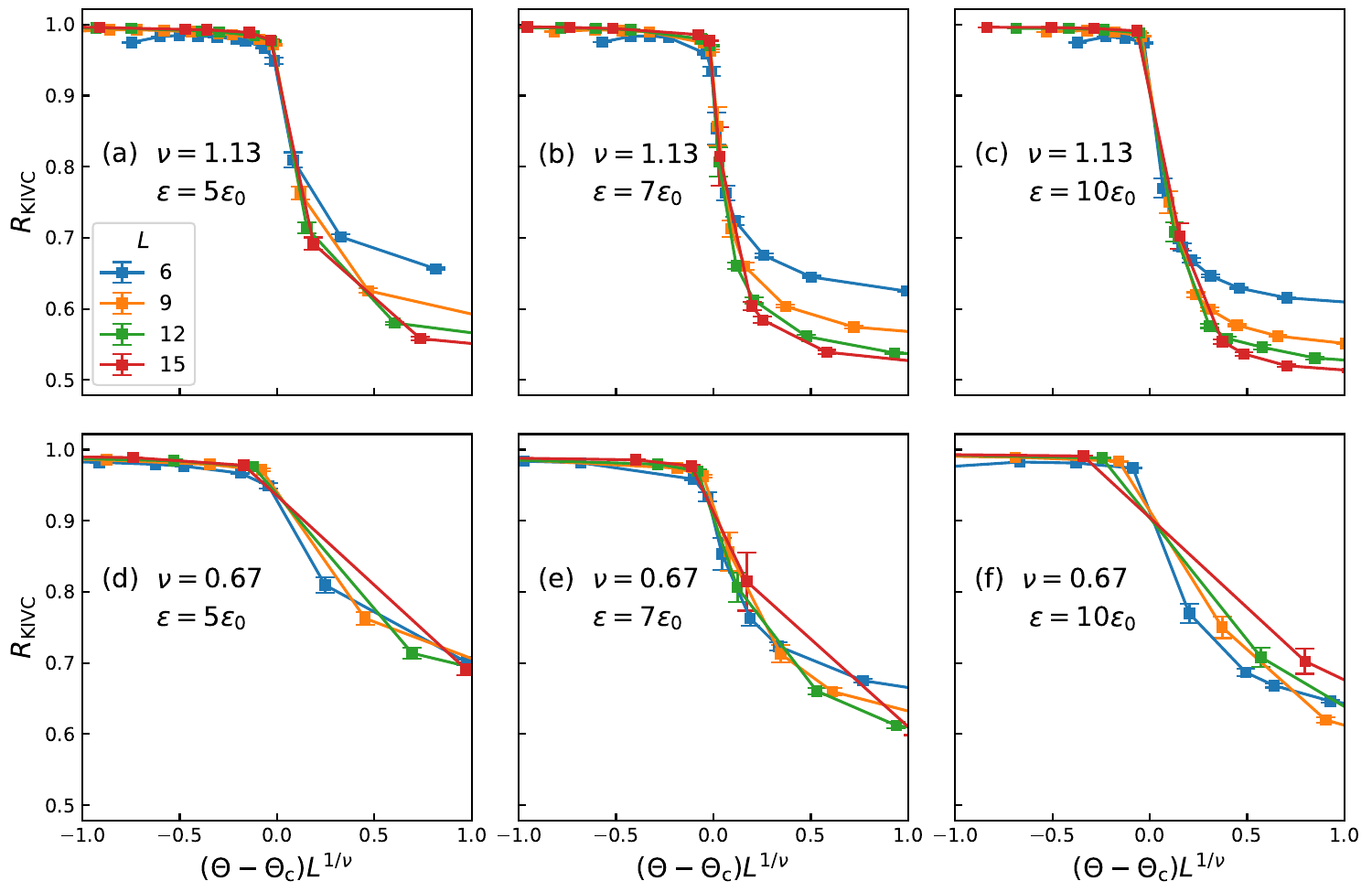}
\caption{\textbf{Data collapses of KIVC correlation with the chiral XY
GN-QCP exponent $\nu = 1.13$ and the (2+1)D XY exponent $\nu = 0.67$.} Here $u_0 = $ 60 meV while each column is with $\varepsilon/\varepsilon_0 = $ 5, 7, and 10 respectively. (a)-(c) are with $\nu = 1.13$, which are the same as the third column of Fig.~\ref{fig:fig4} in the main text, and (d)-(f) are with $\nu = 0.67$.}
\label{fig:fig10}
\end{figure}

Here we show the comparison on the data collapses of the KIVC order parameter with the exponents of the chiral XY GN and (2+1)D XY universalities, as in Fig.~\ref{fig:fig10}. The much less satisfactory data collapses with the (2+1)D XY exponent $\nu = 0.67$ clearly indicate that the (2+1)D XY universality, without the involvement of the Dirac fermions of the GN-QCP, cannot faithfully describe the data and the transitions.

\section{Comparison with exact groundstate of the interacting Hamiltonian}
Following Ref. \cite{Bernevig2021May}, we calculate the charge $\pm 1$ excitations when the kinetic energy is omitted. Based on the symmetry arguments therein, the ground state  at $\nu=0$, which is the case we consider in this manuscript is exact and given in terms of a product state, $|\Psi\rangle=\prod_{s,\eta,\bk,m}c^\dagger_{s,\eta,\bk,m}|0\rangle$, which satisfies $\delta\rho_\bQ|\Psi\rangle=0$. Therefore,
\begin{equation}
\begin{aligned}
    \left[H_\mathrm{I},c^\dagger_{s,\eta,\bk,m}\right]|\Psi\rangle=&H_\mathrm{I}c^\dagger_{s,\eta,\bk,m}|\Psi\rangle\\
    =&\frac{1}{2\Omega}\sum_nR^{s,\eta}_{mn}(\bk)c^\dagger_{s,\eta,\bk,n}|\Psi\rangle,
\end{aligned}
\end{equation}
Notably, in this limit the ground state forms a degenerate U(4) manifold. We can obtain the charge gaps by diagonalizing \begin{equation}
    R^{s,\eta}_{mn}(\bk)=\sum_{\bQ,m'}V(\bQ){\lambda^{s,\eta}_{m,m'}}^*(\bk,\bQ)\lambda^{s,\eta}_{n,m'}(\bk,\bQ).
    \label{eqn:Rmnk}
\end{equation}

\begin{figure*}[!ht]
\centering
\includegraphics[width=0.8\linewidth]{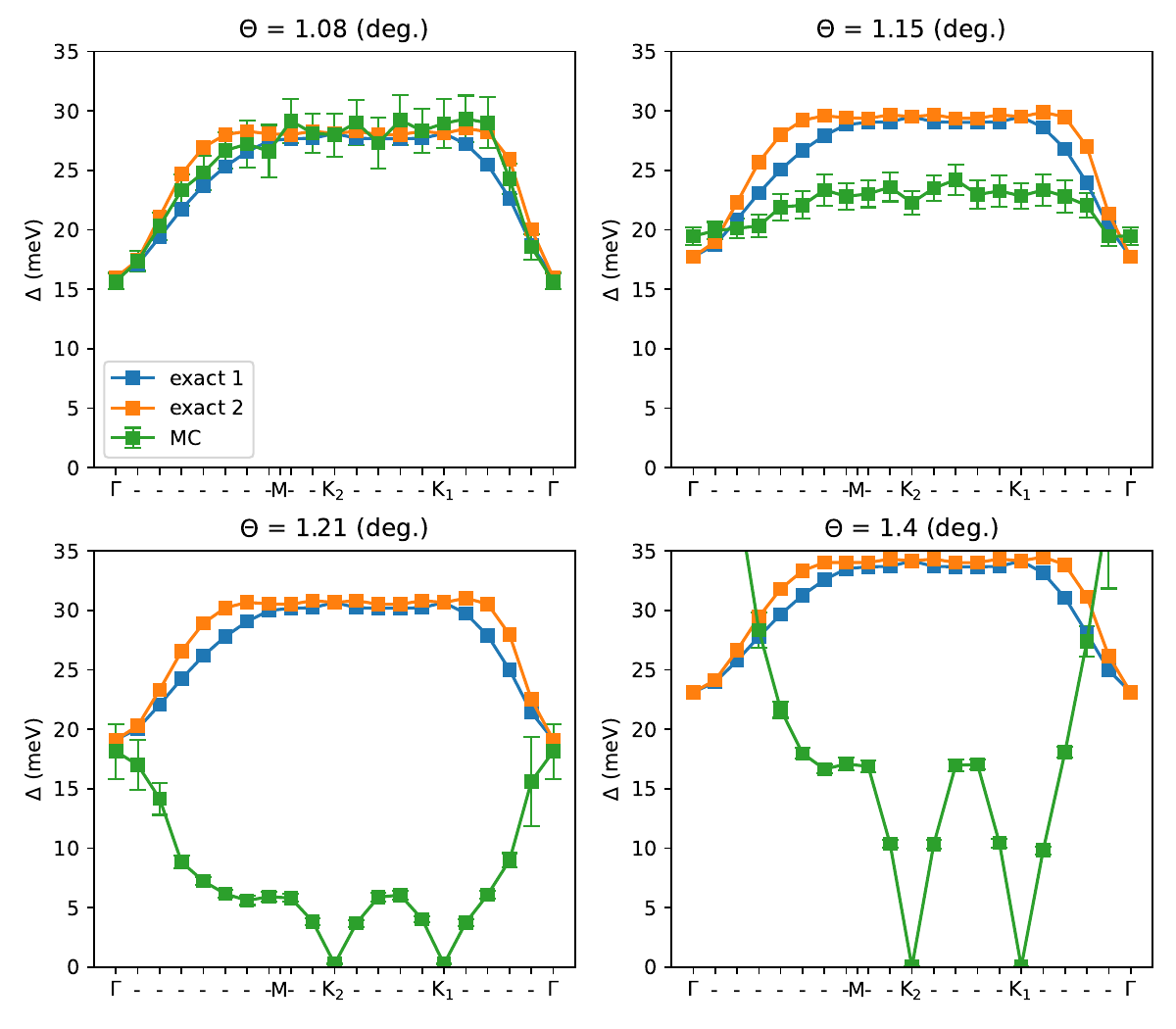}
\caption{Comparison between the gaps obtained from Eq.~\ref{eqn:Rmnk} and from CFMC. This indicates that the non-trivial inclusion of the kinetic energy leads to the gaps closing and the emergence of Dirac cones at $\bK_{1,2}$. Note that the gaps obtained from two methods are comparable at the MA.}
\label{fig:exact}
\end{figure*}

The charge gaps from this expression are used to compare with those from our CFMC simulations. We find that the gaps obtained from both methods are almost identical near the MA ($\Theta=1.08\degree$, where the bandwidth of the flat bands becomes minimal and the Coulomb repulsion dominates) as shown in Fig.~\ref{fig:exact}. The inclusion of the kinetic energy splits the U(4) degeneracy and picks out the KIVC state as has also been described in Refs. \cite{hofmannFermionic2022,bultinck2020ground}. Further increasing the angle, we see a considerable deviation and the emergence of Dirac cones at $\bK_{1,2}$ due to the kinetic energy having a sizable contribution.

\section{On data series and autocorrelations by CFMC and DQMC}

\begin{figure}[!h]
\centering
\includegraphics[width=0.5\linewidth]{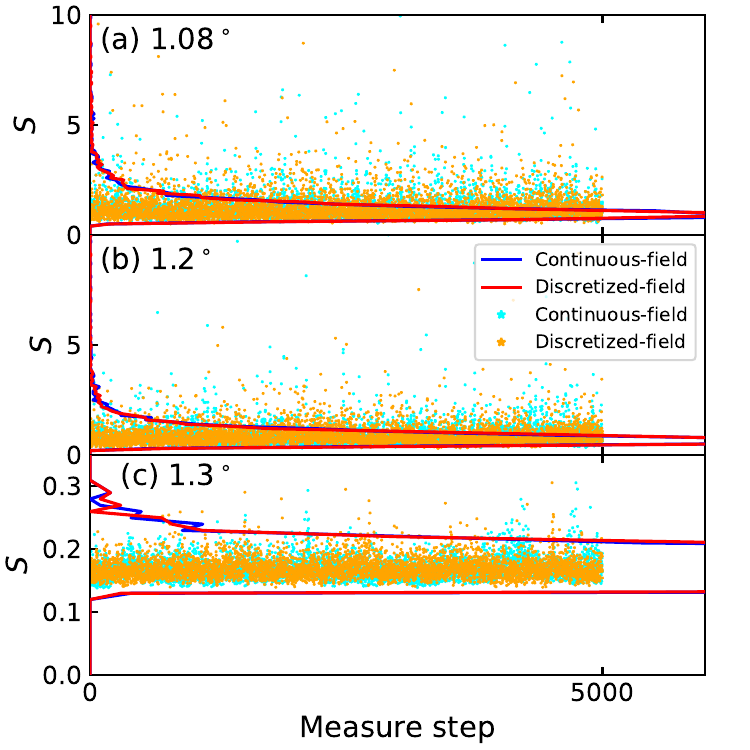}
\caption{\textbf{Time series of an IVC structure factor between DQMC and CFMC.} The time series of the IVC structure facotr for $L=6$ from DQMC and CFMC are as orange stars and cyan stars respectively, and correponding histograms are as red line and blue line respectively, where (a) is of 1.08$\degree$, (b) is of 1.2$\degree$, and (c) is of 1.3$\degree$. Note that the y scales in (c) is different.}
\label{fig:fig12}
\end{figure}

\begin{figure}[!h]
\centering
\includegraphics[width=0.5\linewidth]{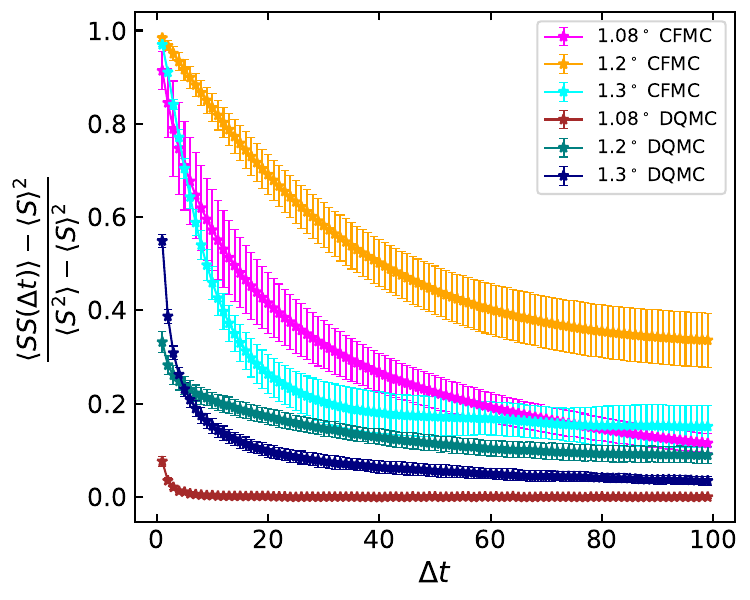}
\caption{\textbf{Autocorrelation for CFMC and DQMC with $L=6$ at different measurement intervals, $\bm{\triangle t}$.} For the simulation at MA, autocorrelation from CFMC at the $\Delta t=100$ interval is comparable with that of DQMC at the $\Delta t=1$ interval. Thus CFMC is only $L^2\times3/100=1.08$ times faster than DQMC in this particular case. However the speedup is much larger for larger $L$, since the autocorrelation is not dependent on lattice size. Thus the speedup is 13.5 times for $L=15$ simulations at MA. The numbers are similar for $1.2\degree$ angle and even more favorable for CFMC at $1.3\degree$ angle. In the latter case, the autocorrelation is the same for CFMC at $\Delta t=17$ interval and DQMC at $\Delta t=1$, which means another factor of 6 for the speedup even for the smallest lattice.}
\label{fig:fig13}
\end{figure}

Comparison of performances of CFMC and discretized-field QMC (DQMC) algorithms is shown in Fig.~\ref{fig:fig7}, Fig.~\ref{fig:fig12}, and Fig.~\ref{fig:fig13} using the same physical settings, on computational time, measured data series, and autocorrelations respectively. The details are narrated in the Methods section of the main text.

\end{document}